\documentclass[]{article}
\usepackage{amsmath}
\usepackage{ifsym}
\usepackage{wasysym}
\usepackage{latexsym}
\usepackage{amssymb}
\usepackage{bbm}
\usepackage{amsthm}
\usepackage{graphicx}

\newcommand{\SU}{\mathrm{SU}}

\newcommand{\rd}[1]{\mathcal{#1}}
\newcommand{\Ref}[1]{(\ref{#1})}

\newcommand{\lalg}[1]{\mathfrak{#1}}  

\newcommand{\eqa}{\begin{eqnarray}}
\newcommand{\neqa}{\end{eqnarray}}
\newcommand{\be}{\begin{equation}}
\newcommand{\ee}{\end{equation}}

\newcommand{\ket}[1]{|{#1}\ra}
\def\ra{\rangle}

\theoremstyle{plain}
\newtheorem{theorem}{Theorem}
\newtheorem{proposition}[theorem]{Proposition}
\newtheorem{lemma}[theorem]{Lemma}

\theoremstyle{definition}

\theoremstyle{remark}

\theoremstyle{plain}
\newcommand{\startproof}{\textbf{Proof.}}
\newcommand{\finishproof}{\hfill $\Box$ \\}

\newcommand{\half}{\frac{1}{2}}
\newcommand{\mspan}{\mathrm{span}}
\newcommand{\dif}{\mathrm{d}}
\newcommand{\deriv}[2]{\frac{\dif #1}{\dif #2}}

\newcommand{\double}[2]{\hspace{0.1em} #1 \hspace{#2} #1 \hspace{0.1em}}
\newcommand{\dbldif}{\double{\dif}{-0.45em}}

\newcommand{\dblwedge}{\double{\wedge}{-0.78em}}

\newcommand{\Z}{\mathbb{Z}}

\newcommand{\Tr}{\mathrm{Tr}}
\newcommand{\SO}{\mathrm{SO}}
\newcommand{\so}{\mathfrak{so}}

\newcommand{\Hil}{\mathcal{H}}

\begin{document}
\title{{\LARGE \bf Flipped spinfoam vertex and loop gravity}}
\author{Jonathan Engle, Roberto Pereira and Carlo Rovelli
 \\[1mm]
\normalsize \em CPT%
\footnote{Unit\'e mixte de recherche (UMR 6207) du CNRS et des Universit\'es
de Provence (Aix-Marseille I), de la Meditarran\'ee (Aix-Marseille II) et du Sud (Toulon-Var); laboratoire affili\'e \`a la FRUMAM (FR 2291).} , CNRS Case 907, Universit\'e de la M\'editerran\'ee, F-13288 Marseille, EU}
\date{\small\today}
\maketitle\vspace{-7mm}
\begin{abstract}

\noindent
We introduce a vertex amplitude for 4d loop quantum gravity. We derive it from a
conventional quantization of a Regge discretization of euclidean general relativity.
This yields a spinfoam sum that corrects some difficulties of the
Barrett-Crane theory.  The second class simplicity constraints are 
imposed \emph{weakly}, and not \emph{strongly} as in Barrett-Crane
theory. Thanks to a flip in the 
quantum algebra, the boundary states turn out to match those of 
$SO(3)$ loop quantum gravity -- the two can be identified as
eigenstates of the same physical quantities -- providing a solution to the
problem of connecting the covariant $SO(4)$ spinfoam formalism
with the canonical $SO(3)$ spin-network one. The vertex amplitude is $SO(3)$ and 
$SO(4)$-covariant.  It rectifies the triviality of the intertwiner
dependence of the  Barrett-Crane vertex, which is responsible for its failure
to yield the correct propagator tensorial structure.
 The construction provides also an
independent derivation of the kinematics of loop quantum gravity
and of the result that geometry is quantized.

\end{abstract}

\section{Introduction}

While the \emph{kinematics} of loop quantum gravity (LQG) \cite{lqg}
is rather well understood \cite{alrev,libro},
its \emph{dynamics} is not understood as cleanly. Dynamics is
studied along two lines: hamiltonian \cite{thomas} or covariant. The
key object that defines the dynamics in the covariant language is
the vertex amplitude, like the vertex amplitude  $\  \sim\!\sim
\put(-2.5,2.5){\circle*{4}} \!\!\!\!<  \  \ = e \gamma^\mu
\delta(p_1+p_2+k) $ that defines the dynamics of perturbative QED.
What is the vertex of LQG?

The spinfoam formalism \cite{spinfoam,spinfoam11,BC1,BC2,spinfoams} can 
be viewed as a
tool for answering this question: the spinfoam vertex plays
a role similar to the vertices of Feynman's covariant quantum field theory.
This picture is nicely implemented in three
dimensions (3d) by the Ponzano-Regge model \cite{PoRe}, whose
boundary states match those of LQG \cite{PRc} and whose
vertex amplitude can be obtained as
a matrix element of the hamiltonian of 3d LQG \cite{KaPe}. But  the
picture has never been fully
implemented in 4d.   The best studied model in the 4d euclidean context is the
Barrett-Crane (BC) theory \cite{BC2}, which is based on the
vertex amplitude introduced by Barrett and Crane \cite{BC1}.  This 
is simple and elegant, has remarkable
finiteness properties \cite{finiteness}, but the suspicion that something is wrong with
it has long been agitated.  Its boundary state space is
similar to, but does not exactly match, that of LQG;
in particular the volume operator is ill-defined.  Worse,
recent results  \cite{emanuele} indicate that it appear to fail to
yield the correct tensorial structure of the graviton propagator 
in the low-energy limit \cite{gravprop}.

It is then natural to try to correct the BC model \cite{mike1,Sergei,se1}.  The
difficulties are all related to the fact that in the BC model the \emph{intertwiner}
quantum numbers are fully constrained. This follows from the fact that
the simplicity constraints are imposed as strong operator
equations ($C_n \psi=0$). However, these constraints are second class
and it is well known that imposing second class constraints strongly may lead to the incorrect
elimination of physical degrees of freedom \cite{dirac}.
In this paper we show that the simplicity constraints can be imposed
weakly ($\langle\phi \,C_n\, \psi\rangle=0$), and that the resulting theory
has remarkable features. First, its boundary quantum state space matches
\emph{exactly} the one of $SO(3)$ LQG: no degrees of freedom
are lost. Second, as the degrees of freedom missing in BC are recovered, the
vertex may yield the correct low-energy $n$-point functions.  Third, the vertex
can be seen as a vertex over $SO(3)$ spin networks or $SO(4)$ spin networks,
and is both $SO(3)$ and $SO(4)$ covariant.

These results have been anticipated in a letter \cite{epr1}. Here we
derive them via a proper quantization of a discretization of
euclidean general relativity (GR).   Indeed, although spinfoam
models have been derived in a number of different manners
\cite{spinfoam11,BC1,BC2,Perez}, most derivations involve peculiar procedures
or intuitive and ad hoc steps.  It is hard to find a proper
derivation of a spinfoam model from the classical field theory,
which follows well-tested quantization procedures.  Here we try to
fill this gap.

From the experience with QCD, one can derive the persuasion that a
nontrivial quantum field theory should be related to a natural
lattice discretization of the corresponding classical field theory
\cite{Immirzi:1996dr}. This persuasion is reinforced by the LQG
prediction of an actual physical discretization of spacetime. Here,
we reconstruct the basis of the euclidean spinfoam formalism as a
proper quantization of a \emph{lattice} discretization of GR.
Conventional lattice formalisms as the ones used in QCD are not very
natural for GR, since they presuppose a background metric. Regge has
found a particularly natural way to discretize GR on a lattice
\cite{regge}, known as Regge calculus.  Quantization of Regge calculus has
been considered in the past  \cite{Rocek:1982fr} and its relation 
to spinfoam theory has been pointed out (see \cite{ruth} and
references therein). Here we express Regge calculus in terms of the
elementary fields used in the loop and spinfoam approach, namely
holonomies and the Plebanski two-form, and we study the quantization
of the resulting discrete theory (on lattice derivations of loop gravity,
see \cite{spinfoam11,Zapata:2004xq,Dass:2006sp}).

The technical ingredient that allows a nontrivial intertwiner state 
space to emerge can be interpreted as a ``flip" of the $SO(4)$ 
algebra, namely an opposite choice of sign in one of its two $SU(2)$ 
factors. The possibility and the relevance of a ``flipped" symplectic 
structure was noticed by Baez and Barrett in \cite{baezbarrett} (where 
they attribute the observation to Jos\'e-Antonio Zapata) and by 
Montesinos \cite{Montesinos:2006qg}.
Montesinos, in particular, has discussed the classical indeterminacy of the 
symplectic structure in detail.  The flip can be viewed as the equivalent 
of the Ashtekar ``trick", which yields a connection as phase space 
variable. The $SO(4)$ generators turn out to directly correspond to 
the bivectors associated to the Regge triangles, rather than to their 
dual.  Using this, we find a nontrivial subspace of the $SO(4)$ 
intertwiner space, which corresponds to closed tetrahedra and  
maps naturally to an $SO(3)$ intertwiner space.

This path leads to a quantum theory that appears to improve several 
aspects of the better known spinfoam models.  In particular: (i)  the 
geometrical interpretation for all the variables becomes fully 
transparent; (ii) the boundary states fully capture the gravitational 
field boundary variables; and (iii)  correspond \emph{precisely} to 
the spin network states of LQG.  The identification is not arbitrary: the 
boundary states of the model are precisely eigenstates of the same 
quantities as the corresponding LQG states. This last result provides 
a solution to the long-standing difficulty of connecting the covariant 
$SO(4)$ spinfoam formalism with the $SO(3)$ canonical LQG one. 
It also provides a novel independent derivation of the LQG kinematics, 
and, in particular, of the quantization of area and volume.  Finally, (iv) 
the vertex of the theory is similar to, but different from, the BC vertex, 
leading to a dynamics that might be better behaved in the low-energy limit.

The paper is organized pedagogically and is largely self contained.
Section 2 reviews background material:  properties
of $SO(4)$ and its selfdual/anti-selfdual split, and the
definition of the fields and the formulation of classical GR as a
constrained Plebanski theory. In section 3 we \emph{discretize} the
theory on a fixed triangulation of spacetime. We do so simply by
taking standard Regge calculus and re-expressing it in terms of the
(discretized) Plebanski two-form. The resulting theory
is governed by the geometry of a 4-simplex, which we
illustrate in detail.  All basic relations among the variables have
a simple interpretation in these terms. The 10 components of the
metric tensor $g_{ab}$ in a point (or in a cell) can be interpreted
as a way to code the 10 variables determining the shape of the cell.
In particular, the norms of the discretized $B$ fields
on the faces are their areas and the scalar product on adjacent
triangles codes the angle between the triangles. While these
``angles" and ``areas" are independent in BF theory, they are
related if they derive from a common metric, namely in GR. In
Section 4 we study the quantization of the system.
We explain the difficulties of imposing the constraints strongly, study
the weak constraints and write their solution. Finally we construct the
vertex amplitude.

We work in the euclidean signature, and on a fixed triangulation.
The issues raised by recovering triangulation independence and the
relation with the Lorentzian--signature theory will be  discussed
elsewhere.

\section{Preliminaries: Plebanski two-form and structure of $SO(4)$}

Riemannian general relativity (GR) is defined by a riemannian metric
$g_{ab}(x)$, where $a,b=1,2,3,4$ and the Einstein-Hilbert  action
\be S[g] = \int \sqrt{g}\  R = \int \sqrt{g}\, g^{ab}R_{ab}.
\label{actiongr} \ee where $g^{ab}$ is the inverse, $g$ the
determinant, and $R_{ab}$ the Ricci curvature of  $g_{ab}$. A good
number of reasons, such as for instance the fact that this metric
formulation is incompatible with the coupling with fermions, suggest
to use the tetrad field $e^a_I(x), I=1,2,3,0$, (the value 0 instead
of 4 is for later convenience and does not indicate a Lorentzian
metric) or its inverse,
namely the tetrad one-form field $e^I(x)=e^I_a(x)dx^a$, to describe
the gravitational field.  This is related to the metric by
$e_a^Ie_b^I=g_{ab}$.  Sum over repeated indices is understood,
and the up or down position of the  $I$ indices is irrelevant. The
spin connection of the tetrad field is an $SO(4)$ connection
$\omega^{IJ}[e]$ satisfying the torsion-free condition \be De^I=
de^I + \omega^I{}_J[e]\wedge e^J = 0. \label{torsionfree} \ee The
action (\ref{actiongr}) can be rewritten as a function of $e^I$ in
the form
\begin{equation}
S[e] = \int  (\det{e})\, e^a_I e^b_J F_{ab}^{IJ}[\omega[e]] =
\frac{1}{2}\int \epsilon_{IJKL}\ e^I \wedge e^J \wedge
F^{KL}[\omega[e]]
\label{action2}
\end{equation}
where $F[\omega]$ is the curvature of $\omega$. Alternatively, GR
can be defined in first order form in terms of independent variables
$\omega^{IJ}$ and $e^{I}$, by the action \be S[e,\omega]  =
\frac{1}{2}\int \epsilon_{IJKL}\ e^I \wedge e^J \wedge
F^{KL}[\omega]. \label{actiongr_1stOrd} \ee In this case, \Ref{torsionfree}
is obtained as the equation of motion for $\omega$.

\subsection{Plebanski two-form and simplicity constraints}

At the basis of the spinfoam formalism is the use of the Plebanski
two-form $\Sigma^{IJ} \equiv \frac{1}{2}\Sigma^{IJ}_{ab}\ dx^a\wedge dx^b$,
defined as
\be
\Sigma^{IJ} =  e^I \wedge e^J. \label{Sigma}
\ee
or its dual, usually called $B^{IJ} \equiv 
\frac{1}{2}B^{IJ}_{ab}\ dx^a\wedge dx^b$ for 
a reason that will be clear in a moment, defined as
\be
B^{IJ}=  \frac{1}{2}\epsilon^{IJ}{}_{KL}\  \Sigma^{KL}
 =  \frac{1}{2}\epsilon^{IJ}{}_{KL}\   e^K \wedge e^L \label{Bee}
\ee
We use the following
notation for two-index objects: a scalar product: $A \cdot B\equiv
A^{IJ} B_{IJ}$; a norm:  $|B|^2 \equiv B \cdot B$, and the duality
operation $({}^*B)^{IJ} = \frac12 \epsilon^{IJ}{}_{KL} \ B^{KL} $.
So that, in particular, ${}^*B\cdot B = \frac12 \epsilon_{IJKL} \
B^{IJ}B^{KL} $.  Thus we write \Ref{Bee} in the form
\be
B = {}^*\Sigma =  *(e^I\wedge e^J)
\ee
The geometrical interpretation of the Plebanski
two-form (or the $B$ two-form) is captured by the following. Observe that
\be
       |\Sigma_{ab}|^2 =  |B_{ab}|^2 =g_{aa}g_{bb}- g_{ab}g_{ab} \equiv  2A^2_{ab}
\label{areael}
\ee
and
\be
\Sigma_{ab}\cdot \Sigma_{ac}=       B_{ab}\cdot B_{ac}=
g_{aa}g_{bc}- g_{ab}g_{ac} \equiv  2J_{aabc}
\ee
The quantity $A_{ab}$ gives the \emph{area} element 
$A_{ab}dx^ady^b$ of the infinitesimal surface $dx^ady^b$.  
 Therefore we can write
\be
\int_S |\Sigma| = \int_S |B| \ \ \equiv \  \int_S |\Sigma_{ab}|\ dx^a dy^b
= \sqrt{2} \times Area(S)
\ee
The quantity $J_{aabc}$ is the related to the \emph{angle}  $\theta_{aabc}$ 
between the surface elements $dx^ady^b$ and $dx^adz^c$. In fact, if we 
take the scalar product of the normals of these two surface elements (in 
the 3-space they span), we obtain (without writing the infinitesimals vectors)
\be
A_{ab}A_{ac}\cos\theta_{aa\; bc} =
g^{ef} (\epsilon_{egh}\delta^g_a\delta^h_b)(\epsilon_{fgh}\delta^g_a\delta^h_c)
=g_{aa}g_{bc}-g_{ab}g_{ac} = J_{aabc}.
\ee
Finally, the 4-form
\be
V \equiv
  \frac{1}{4!}\epsilon_{IJKL} \Sigma^{IJ}\wedge \Sigma^{KL}
=  \frac{1}{4!}\epsilon_{IJKL} B^{IJ}\wedge B^{KL}
\ee
is easily seen to be (proportional to) the volume element $dV= \sqrt{g} \ d^4x$. Intuitively, describing the geometry in terms of $\Sigma$ rather than $g_{ab}$ is using  as elementary variable areas and angles rather than length and angles.

Using the Plebanski field, the action can be written in the BF-like form
\be
S[e,\omega] =
\frac12\int \epsilon_{IJKL}\ \Sigma^{IJ}[e] \wedge F^{KL}[\omega]
= \int B_{IJ}[e] \wedge F^{IJ}[\omega].
\label{BFaction}
\ee
The reason this action defines GR and not BF theory is that the independent 
variable to vary is the tetrad $e$, not the two-form $B$.  While the BF field 
equations are obtained by varying the action  (\ref{BFaction}) under 
arbitrary variations of $B$ (and $\omega$), GR is defined by varying 
this action under the variations that respect the form (\ref{Bee}) of the 
field $B$.  This condition can be expressed as a constraint equation 
for $\Sigma$:
\be
\Sigma^{IJ} \wedge \Sigma^{KL}= V\ \epsilon^{IJKL}
\label{Pleb1}
\ee
Equivalently ,
\be
 {}^*\Sigma_{ab}\cdot  \Sigma_{cd}= 2\tilde{V}\ \epsilon_{abcd}.
\label{Plebb}
\ee
where $V=\frac{1}{4!}\tilde{V}\epsilon_{abcd}dx^a \wedge ... \wedge dx^d$.
This system of constraint can be decomposed in three parts:
\begin{eqnarray}
a)\ \ \ \ \   {}^*\Sigma_{ab}\cdot  \Sigma_{ab}&=&0\ ,\label{a}\\
b)\ \ \ \ \   {}^*\Sigma_{ab}\cdot  \Sigma_{ac}&=&0\   ,\label{b}\\
c)\ \ \ \ \   {}^*\Sigma_{ab}\cdot  \Sigma_{cd}&=&\pm 2\tilde V\ .
\label{c}
\end{eqnarray}
where the indices $abcd$ are all different, and the sign in the last
equation is determined by the sign of their permutation. Equivalently,
the $B$ field satisfies these same equations. These three
constraint play an important role in the following. They are called
the \emph{simplicity} constraints.  GR can be written as an $SO(4)$
BF theory whose $B$ field satisfies the simplicity constraints
(\ref{a}-\ref{b}-\ref{c}).

\subsection{Selfdual structure of $SO(4)$}\label{selfdual}

In this section we recall some elementary facts about $SO(4)$ and we
make an observation about its representations that plays a role in the
following.

The group $SO(4)$ is
locally isomorphic to the product of two subgroups, each loc. isomorphic
to $SU(2)$: $SO(4)\sim \left( SU(2)_+\times SU(2)_- \right)/
\mathbb{Z}_2$. That is, we can write each $U\in SO(4)$ in the form
$U=(g_+,g_-)$ where $g_+\in SU(2)_+$ and $g_-\in SU(2)_-$ and
$UU'=(g_+g'_+,g_-g'_-)$. This is clearly seen looking at its algebra
$\lalg{so(4)}$, which is the linear sum of two commuting
$\lalg{su(2)}$ algebras. Explicitly, let $J^{IJ}$ be the generators
of $\lalg{so(4)}$. Define the selfdual and anti-selfdual generators
$J_\pm :=  *J \pm J, $ that satisfy $J_\pm =\pm  *J_\pm$. Then it is
immediate to see that $[J_+,J_- ]=0$. The $J_+$ span a three
dimensional subalgebra $\lalg{su(2)}_+$ of $\lalg{so(4)}$, and the
$J_-$ span a three dimensional subalgebra $\lalg{su(2)}_-$ of
$\lalg{so(4)}$, both isomorphic to $\lalg{su(2)}$.

It is convenient to choose a basis in $\lalg{su(2)}_+$ and in
$\lalg{su(2)}_-$.   For this, choose a unit norm vector $n$ in
$R^4$, and three other vectors $v_i,
i=1,2,3$ forming, together with $n$, an orthonormal basis, for
instance $v_i^I=\delta^I_i$, and define
\begin{equation}
J_\pm^i = \half({}^*J\pm J)_{IJ}\ v_i^I n^J
\end{equation}
The $\lalg{su(2)}$ structure is then easy to see, since
$[J_\pm^i,J_\pm^j]= \epsilon^i{}_{jk} J_\pm^{k} $.
In particular, we can choose $n=(0,0,0,1)$, and $v_i^I=\delta^I_i$, and we have
\begin{equation}
J_\pm^i = -\frac{1}{4}
\epsilon^i{}_{jk}J^{jk}  \pm \half J^{i0} .
\label{basis}
\end{equation}

Notice that in choosing this basis we have broken $SO(4)$
invariance. In fact, the split $\lalg{so(4)}=\lalg{su(2)}_+\oplus
\lalg{su(2)}_-$ is canonical, but there is no canonical isomorphism
between $\lalg{su(2)}_+$ and $\lalg{su(2)}_-$ or between $SU(2)_+$
and $SU(2)_-$. One such isomorphism $G_n:  SU(2)_+ \to SU(2)_-$ is
picked up, for instance, by choosing the vector $n$. It sends
$g_+$ to $g_-$, where the element $(g_+,g_-)$ of $SO(4)$ leaves $n$
invariant. This isomorphism defines a notion of \emph{diagonal}
elements of the $\lalg{so(4)}$ algebra: the ones of the form $a_i
J^i_+ + a_i J^i_-$. Exponentiating these, we get the \emph{diagonal}
elements of the $SO(4)$ group, which we can write as $U=(g,g)$.
These diagonal elements form an $SU(2)$ subgroup of $SO(4)$, which
is \emph{not} canonical: it depends on $n$.  It is the subgroup of
$SO(4)$ that leaves the vector $n$ invariant. If we consider the 3d
surface (``space") orthogonal to $n$, the diagonal part of $SO(4)$
is (the double covering of) the $SO(3)$ group of the (``spatial")
rotations of this space; we denote it $SO(3)_n\subset SO(4)$.
Borrowing from the Lorentzian terminology, we can call ``boost" a
change of $n$. Its effect is to rotate the $\pm$ bases relative to
one another.

Notice that for any two-index quantity $B^{IJ}$,
\be
          \frac{1}{4}B\cdot B = B_+^i B_+^i +  B_-^i B_-^i,
\ee
while
\be
          \frac{1}{4}{}^*B\cdot B = B_+^i B_+^i -  B_-^i B_-^i.
          \label{pippo}
\ee This split \emph{is} independent from $n$, as the norms are not
affected by a rotation of the basis. In particular, $C = \frac{1}{4}
J\cdot J$ and $\tilde C = \frac{1}{4} {}^*J\cdot J$ are the scalar
and pseudo-scalar quadratic Casimirs of $\lalg{so(4)}$. They are,
respectively, the sum and the difference of the quadratic Casimirs
of $\lalg{su(2)}_+$ and $\lalg{su(2)}_-$. The representations of the
universal cover of $SO(4)$, the group $Spin(4) \sim SU(2) \times
SU(2)$ are labelled by two half integers $(j_+ ,j_-)$. The
representations of $SO(4)$ form the subset of these for which $j_+ +
j_-$ is integer. The representations satisfying $j_+ =j_-$, which
clearly belong to this subset, are called \emph{simple}:  they play
a major role in the BC theory as well as in the quantization below.

The following observation plays a major role in section
\ref{intspace_sect}. Consider a simple representation $(j,j)$. This
is also a representation of the subgroup $SO(3)_n\subset SO(4)$, but
a reducible one. Clearly, it transforms in the representation
$j\otimes j$, where $j$ indicates the usual spin $j$ representation
of $SU(2)$. If we decompose it into irreducible representations of
$SO(3)_n$, we have \be
      (j,j) \ \ \rightarrow \ \  j\otimes j \ \  = \ \  0 \oplus 1 \oplus ...  \oplus (2j-1) \oplus 2j.
\ee
The value of $C_4$ in this representation is $2(j(j+1))$. Consider the value
that the Casimir $C_3$ of the subgroup $SO(3)_n$ takes on the lowest and
highest-spin representations.  $C_3$ vanishes on the spin-0 representation.
On the spin-$2j$ representation, it
has the value $C_3= 2j(2j+1)$, which is related to $C_4$ by
\begin{equation}
\sqrt{C_3+1/{4}}-\sqrt{2C_4+1}+1/{2}=0
\label{qcas1}
\end{equation}
and in the large $j$ limit by
\begin{equation}
C_3=2C_4,
\end{equation}
that is, if we have chosen $n=(0,0,0,1)$,
\begin{equation}
J^{IJ}J_{IJ}=J^{ij}J_{ij}+2J^{0i}J_{0i},
\end{equation}
which implies
$J^{0i}=0$. Therefore, the spin-zero and the spin-$2j$ components of the
simple $SO(4)$ representation $(j,j)$ are
characterized  respectively by
\begin{eqnarray}
{\rm spin}\ 0&:&\ \  J^{ij}=0,  \nonumber \\
{\rm spin}\ 2j&:&\ \  J^{i0}=0 \label{zumzum}
\end{eqnarray}
in the ``classical" large-$j$ limit.

\section{Regge discretization}

We now approximate euclidean GR by means of a
discrete lattice theory.  A very natural way of doing so is Regge
calculus  \cite{regge}.  The idea of Regge calculus is the following.
The object described by euclidean GR is a Riemannian manifold
$(M,d_M)$, where $M$ is a differential manifold and $d_M$ its
metric.  A Riemannian manifold can be approximated by means of a
\emph{piecewise flat} manifold $(\Delta, d_\Delta)$, formed by flat
(metric) simplices (triangles in 2d, tetrahedra in 3d, 4-simplices
in 4d...) glued together in such a way that the geometry of their
shared boundaries matches. Here $\Delta$ is the abstract
triangulation and $d_\Delta$ is its metric, which is determined by
the size of the individual simplices. For instance, a curved 2d
surface can be approximated by a surface obtained by gluing together
flat triangles along their sides: curvature is then concentrated on
the points where triangles meet, possibly forming ``the top of a
hill".    With a sufficient number $N$ of simplices, we can
approximate sufficiently well any given (compact) Riemannian
manifold $(M, d_M)$, with a Regge triangulation $(\Delta,
d_\Delta)$.\footnote{For instance, in the sense that the two can be
mapped into each other, $P:M\to \Delta$, in such a way that the
difference between the distances between any two points,
$d_M(x,y)-d_\Delta(P(x),P(y))$,  can be made arbitrary small with
$N$ sufficiently large.}

If we fix the abstract triangulation $\Delta$ and we vary
$d_\Delta$, namely the size of the individual $n$-simplices, then we
can approximate to a certain degree a \emph{subset} of GR fields.
Therefore by fixing $\Delta$ we capture a subspace of the
full set of all possible gravitational fields.  Thus, over a fixed
$\Delta$ we can define an approximation of GR, in a manner analogous
to the way a given Wilson lattice defines an approximation to
Yang-Mills field theory, or  the approximation of a partial
differential equation with finite--differences defines a
discretization of the equation. Therefore the Regge theory over
a fixed $\Delta$ defines a cut-off version of GR.

It is important to notice, however, that the Regge cut-off is neither ultraviolet nor
infrared.  This is sharp contrast with the case of lattice QCD.
In lattice QCD, the number $N$ of elementary cells of the lattice defines
an infrared cut-off: long wavelength degrees of freedom are recovered
by increasing $N$. On the other hand, the physical size $a$
of the individual cells enters the action of the theory, and short wavelength
degrees of freedom are recovered in lattice QCD by decreasing $a$. Hence $a$ is
ultraviolet cut-off.  In
Regge GR, on the contrary, there is no fixed background size of the
cells that enters the action. A fixed $\Delta$ can carry both a very large
or a very small geometry.  The cut-off implemented by $\Delta$ is therefore
of a different nature than the one of lattice QCD.  It is not difficult to see that
it is a cut-off in the \emph{ratio}
between the smallest allowed wavelength and the overall size of the
spacetime region considered.  Thus, fixing  $\Delta$ is equivalent to
cutting-off the degrees of freedom of GR that have much smaller  wavelength
than the arbitrary size $L$ of the region one considers. Since (as we shall see,
and as implied by LQG) the quantum theory has no degrees of freedom below
the Planck scale, it follows that a Regge approximation is good for $L$ small,
and it is a low-energy approximation for $L$ large.\footnote{Since the expansion
parameter  $\lambda$ used in group field theory \cite{gft} is equivalent to the number
of cells in Regge calculus, this discussion clarifies also the physical
meaning of the group field theory $\lambda$ expansion.}

Consider a 4d triangulation.  This is formed by oriented
4-simplices, tetrahedra, triangles, segments and points. We call
$v,t$ and $f$ respectively the 4-simplices, the tetrahedra and the
triangles of the triangulation decomposition.  The choice of the
letters $v$ and $f$ is dictated by the fact that in the
dual-complex, to which we will later shift, 4-simplices are dual to
\emph{vertices}, and triangles are dual to \emph{faces}.\footnote{For
coherence, we should also call $e$ the tetrahedra, which are dual to the
\emph{edges} of the dual complex; this is indeed  the common convention
\cite{libro}.  But in the present context this
would generate confusion with the notation for the tetrad. We
will shift to the $e$ notation for edges only later on, in the
quantum theory part.} The
metric, we recall, is flat within each 4-simplex $v$.  All the
tetrahedra, triangles and segments are flat (and, respectively,
straight). The geometry induced on a given tetrahedron from the
geometry of the two adjacent 4-simplices is the same.  This structure
will allow us to take the geometry of the {\em tetrahedra} as the
fundamental dynamical object, and interpret the constraints implied
by the fact that five tetrahedra fit into a single four-simplex as
the expression of the  dynamics.  It is this peculiar perspective that
makes the construction below possible.

In $d$ dimensions, a $d-2$ simplex is surrounded by a cyclic
sequence of $d$-simplices, separated by the $d-1$ simplices that
meet at the $d-2$ simplex. This cyclic sequence is called the
\emph{link} of the $d-2$ simplex.  For instance, in  dimension 2, a
point is surrounded by a link of triangles, separated by the edges
that meet at the point; in dimension 3, it is an edge which is
surrounded by a link of tetrahedra, separated by the triangles that
meet at the edge; in dimension 4, which is the case that concerns
us, a triangle $f$ is surrounded by a link of 4-simplices $v_1,...,
v_n$, separated by the tetrahedra that meet at the triangle $f$.

In Regge calculus, curvature is concentrated on the $d-2$ simplices.
In dimension 4, curvature is therefore concentrated on the triangles
$f$. It is generated by the fact that the sum of the dihedral angles
of the 4-simplices in the link around the triangle may be different
from 2$\pi$.   We can always choose Cartesian coordinates covering
one 4-simplex, or two adjacent 4-simplices; but in general there are
no continuous Cartesian coordinates covering the region formed by
all the 4-simplices in the link around a triangle.

The variables used by Regge to describe the geometry $d_\Delta$ of
the triangulation $\Delta$ are given by the set of the lengths of
all the segments of the triangulation.  Here we make a different
choice of variables, which matches more closely what happens in the
spinfoam formalism and in loop gravity.

Consider one tetrahedron $t$ in the triangulation.  Choose a
Cartesian coordinate system $x^a$ covering the tetrahedron. Choose
an orthonormal basis \be e(t)=e_{a}^I(t)\ v_I\ dx^a \ee for each
such tetrahedron $t$. Here $v_I$ is a basis in $R^4$, chosen once
and for all.  This quantity of course transforms covariantly under a
change coordinate system $x^a$, therefore it is intrinsically
defined as a one-form with values in $R^4$, associated to the
tetrahedron.   This will be our first variable, giving a discretized
approximation of the gravitational field.

\begin{figure}
\begin{center}
  \includegraphics[height=6cm]{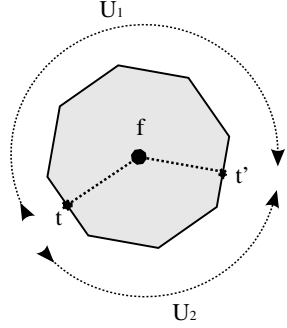}
\end{center}\vspace{-1.6cm}
\caption{The link of the face $f$, in grey, and the two group elements associated to a couple of tetrahedra $t$ and $t'$ in the link.}
\label{twochains}
\end{figure}

Consider the five tetrahedra $t_A, A=1,...5$ that bound a single
simplex $v$.  The five variables $e(t_A)$ are not independent,
because the simplex is flat. Since the simplex is flat, we can
always choose a common Cartesian coordinate system $x^a$ for the
entire $v$. Let $e(v)=e_{a}^I(v)\, v_I\,  dx^a$ denote an
orthonormal basis describing the geometry in $v$. Then each $e(t_A)$
must be related to $e(v)$ by an $SO(4)$ rotation in $R^4$. That is
there must exist five $SO(4)$ matrices $V_{vt_A}$ such that
\begin{equation}
e(v)_a^I = (V_{v t_A})^I{}_J\; e_a^J(t_A)
\end{equation}
in the common coordinate patch. One can also define the
transformation from one tetrahedron to the other:
\begin{equation}
e(t_A) = U_{t_A t_B}(v) e(t_B)
\end{equation}
The later are of course constrained by the former:
\begin{equation}
U_{t_A t_B}(v)=V_{t_A v}\; V_{v t_B}
\label{flat}
\end{equation}
where $V_{t_A v}\equiv V_{vt_A}^{-1}$. Now consider two tetrahedra
$t$ and $t'$ sharing a face but not necessarily at the same four
simplex and define as before the $SO(4)$ transformation between the
two $U_{t t'}$. Remember that one can choose the same coordinate
system for an open chain of four simplices linking two tetrahedra
around the face they share. This constrains $U_{t t'}$ to be of the
form:
\be
U_{t t'}=V_{t v_1} \ ... \  V_{v_n t'}
\label{flatbound}
\ee
where $v_1 \ ... \ v_n$ is the (open) chain of 4-simplices between
$t$ and $t'$ around their common face. Now one must be cautious. For
two arbitrary tetrahedra sharing a face in the interior of the
discretization, there are two such chains (see Figure \ref{twochains})
To resolve this ambiguity, we make use of the orientation of the face
$f$: this orientation gives a notion of clockwise and
counterclockwise around the link.  We choose the convention that
$U_{tt'}$ denote the holonomy around the chain in the clockwise
direction, starting at $t'$ and ending at $t$. (For $f$ on the
boundary, or when one is considering a single $4$-simplex, of course
the orientation of $f$ need not be used to resolve this ambiguity.)

The arbitrariness in the choice of the orthonormal basis at each 
tetrahedron is reflected in the local $SO(4)$ gauge transformations
\begin{eqnarray}
e(t) &\to& \Lambda(t)\; e(t),\\
e(v) &\to& \Lambda(v)\; e(v),\\
V_{vt} &\to& \Lambda(v)\; V_{vt}\; \Lambda(t)^{-1},\\
U_{tt'} &\to& \Lambda(t)\; U_{tt'}\; \Lambda(t')^{-1},
\label{gaugetra}
\end{eqnarray}
where $\Lambda(t),\Lambda(v)\in SO(4)$ are the gauge 
parameters associated with the choice of basis in tetrahedra 
and vertices respectively.

In each tetrahedron $t$, consider the bivector two-form
\be
\Sigma(t)=e(t)\wedge e(t).
\ee
and its dual
\be
B(t)={}^*\Sigma(t).
\ee
In components, this is $B(t)=\half
B^{IJ}_{ab}(t)v_Iv_J dx^a\wedge dx^b$ where \be
B^{IJ}_{ab}(t)=\epsilon^{IJ}{}_{KL}\ e^K_a(t) e^L_b(t). \ee Now,
consider a tetrahedron $t$ and a triangle $f$ in its boundary.
Associate a bivector $\Sigma_f(t)$ to the triangle $f$, as follows.
$\Sigma_f(t)$ is defined as the surface integral of the two-form $\Sigma(t)$
over the triangle
\begin{equation}
\Sigma_f(t)=\int_f \  \Sigma(t). \label{BIJ}
\end{equation}
This is the geometrical bivector naturally associated to the (oriented)
triangle, in the frame associated to the tetrahedron $t$.  Its dual bivector
is $B_f(t)={}^*\Sigma_f(t)$.
A single triangle $f$ belongs to (the boundary of) the several tetrahedra
$t_1, t_2, t_3, ...$ that are around its link.  The bivectors
$\Sigma^{IJ}_f(t_1), \Sigma^{IJ}_f(t_2), \Sigma^{IJ}_f(t_3), ...$ are different,
because they represent the triangle in the internal frames
associated to distinct tetrahedra. But they are of course related by
\begin{equation}
\Sigma^{IJ}_f(t_1) = U_{t_1t_2}{}^I{}_K(v_{12})\;
U_{t_1t_2}{}^J{}_L(v_{12}) \ \Sigma^{KL}_f(t_2). \label{issa}
\end{equation}

Notice that if the tetrad is Euclidean $e^I_a(t)=\delta^I_a$
and the triangle is in the (1,2) plane, the only non-vanishing components
of $\Sigma^{IJ}_f(t)$ are $\Sigma^{12}_f(t)=-\Sigma^{21}_f(t)$, while 
the only non-vanishing components of $B^{IJ}_f(t)$ are $B^{30}_f(t)=-B^{03}_f(t)$.
More in general, if the normal of the tetrahedron is $n^I$, then for all the faces
$f$ of the tetrahedron $t$ we have
\begin{eqnarray}
\epsilon_{IJKL}\ n^J\ B^{KL}_f(t) &=& 0,\\
 n_I \  \Sigma^{IJ}_f(t)&=&0.  \label{zucchero1}
\end{eqnarray}
In particular, if we choose a gauge where $n^I=(0,0,0,1)$, we have
\begin{eqnarray}
B^{ij}_f(t) &=& 0,\\
 \Sigma^{i0}_f(t)&=&0.  \label{zucchero2}
\end{eqnarray}

The quantities $\Sigma^{IJ}_f(t)$ and $V_{vt}$ can be taken as a
discretization of the continuous gravitational fields $\Sigma^{IJ}$ and
$\omega^{IJ}$ that define GR in the Plebanski formalism.

\subsection{Constraints on $\Sigma$}

Since our aim is to promote $\Sigma^{IJ}_{f}(t)$ to an independent variable, let us
study the constraints it satisfies.
It is immediate to see that the four bivectors
$\Sigma^{IJ}_{f_1}(t),...,\Sigma^{IJ}_{f_4}(t)$ associated to the four faces
of a single tetrahedron satisfy the closure relation
\be
\Sigma^{IJ}_{f_1}(t)+\Sigma^{IJ}_{f_2}(t)+\Sigma^{IJ}_{f_3}(t)+\Sigma^{IJ}_{f_4}(t)=0.
\label{closurev}
\ee

Since the two-form $\Sigma^{IJ}(t)$ is defined in terms of a tetrad field, 
it satisfies the relation \Ref{Plebb}, or, equivalently, (\ref{a},\ref{b},\ref{c}).  
Multiplying these relations by the coordinate bivectors representing 
triangles, gives the following relations.  For each triangle $f$, we have
\begin{equation}
*\Sigma_f(t) \cdot \Sigma_f(t)=0.
\label{aa}
\end{equation}
For each couple of adjacent triangles $f,
f'$, we have
\begin{equation}
*\Sigma_f(t) \cdot \Sigma_{f'}(t)=0.
\label{bb}
\end{equation}
We call these two constraints the diagonal and off-diagonal
simplicity constraints, respectively.

The last constraint, for $f$ and $f'$ in the same four simplex sharing just
a point, is a little subtler. It is
\begin{equation}
*\Sigma_f(v) \cdot \Sigma_{f'}(v)= \pm 12 V(v).
\label{ccv}
\end{equation}
where $V(v)$ is the four volume of the simplex and the sign depends
on relative face orientation \cite{Perez}. Of course the volume in
the last formula is irrelevant. What this equation tells us is that
the volume is the same when computed with different choices of pairs
of faces in the same four simplex. Now, using the transformation law
for the bivectors (\ref{issa}) one can write it with tetrads defined
on tetrahedra:
\begin{equation}
*\Sigma_f(t) \cdot \left(U_{tt'}(v)\Sigma_{f'}(t')U_{tt'}^{-1}(v)\right)= \pm 12 V(v).
\label{cc}
\end{equation}
One can show that if (\ref{flat}) is
satisfied and the first set of constraints
(\ref{closurev},\ref{aa},\ref{bb}) is satisfied then (\ref{ccv}) or, equivalently, (\ref{cc})
 is satisfied automatically.

Let's count the degrees of freedom for each four simplex, as a
check. We start with the $60$ degrees of freedom of the bivectors,
$6$ for each face. The constraint (\ref{closurev}) imposes $24$
independent equations. The number of independent equations of the
type (\ref{aa}) are $10$, and finally the constraints of the type
(\ref{bb}) contribute $10$ independent equations. The last
constraint is implicitly imposed when we consider the tetrahedra to
belong to the same four simplex. We are left with the
$60-(24+10+10)=16$ degrees of freedom of the tetrad $e(v)$.

There is however some additional discrete degeneracy.
The set of constraints
(\ref{closurev},\ref{aa},\ref{bb}) has two classes of solutions (see
appendix \ref{app_tetrad}):
\be
\Sigma_{f_1}^{IJ} = 2 e^{[I}_2 e^{J]}_3 \; \text{and cyclically} .
\label{sol1}
\ee
and
\be
\Sigma_{f_1}^{IJ} = \epsilon^{IK}{}_{KL} e^{K}_2 e^{L}_3 \; \text{and cyclically} .
\label{sol2}
\ee
These two sectors of Plebanski theory are well known in the
literature (see \cite{spinfoam},\cite{Perez}); they correspond to the
fact, remarked above, that both $\Sigma$ and $B$ satisfy the
same equations (\ref{closurev},\ref{aa},\ref{bb}).
Because of the double solution \Ref{sol1} and \Ref{sol2}, these
constraints do not determine $\Sigma$ uniquely.

However, we
can give the off-diagonal simplicity constraints a different and
slightly stronger form,
which fixes this degeneracy, and which will play a role below.
We can replace the off-diagonal simplicity constraints with the
following requirement: that for each tetrahedron $t$ there exist
a covariant vector $n_I$ such that \Ref{zucchero1} holds
for all the faces $f$ of the tetrahedron $t$. It is immediate to see that
this implies the off-diagonal simplicity constraints, and that it is
implied by the physical solution \Ref{sol1} of these constraints.
Equivalently, we require that there is a gauge in which
\Ref{zucchero2} holds.
The geometry of this requirement is transparent: $n_I$ is the normal to
the tetrahedron $t$ in four dimensions, and \Ref{zucchero1}, or \Ref{zucchero2},
require  that the faces of the tetrahedron are all confined to the 3d hyperplane
normal to $n_I$.

\subsection{Relation with geometry}

How do we read out the geometry of the riemannian manifold, from the
variables defined? First, it is easy to see that the \emph{area} of
a triangle $f$ is, up to a constant factor, the norm of its
associated bivector: \be \sqrt{2}A_f= |\Sigma_f(t)| \label{area} \ee
Notice that the l.h.s. is independent of $t$.  This is consistent
because the relation between $\Sigma_f^{IJ}(t)$ and $\Sigma_f^{IJ}(t')$ is an
$SO(4)$ transformation, whence the norm is invariant, $ |\Sigma_f(t)|=
|\Sigma_f(t')|$. This shows that the definitions chosen are consistent
with a characteristic requirement of a Regge triangulation: the
boundaries of the flat 4-simplices match, and in particular the area
of a triangle computed from any side is the same.  Of course, it
follows that the same is true for the volume of each tetrahedron.

Given two triangles $f$ and $f'$ in the same tetrahedron $t$, there
is a dihedral angle $\theta_{ff'}$ between the two.  This angle can
be obtained from the product of the normals \be J_{ff'}:= A_f
A_{f'}\cos\theta_{ff'}= \Sigma_f(t)\cdot \Sigma_{f'}(t)/2. \ee
These are also $SO(4)$ invariant, hence well defined independently
from the tetrahedron. The two gauge invariant quantities $A_f$ and
$J_{ff'}$ characterize the geometry entirely.  Notice that we can
view the area as the ``diagonal" part of $J_{ff'}$ and write
$A^2_f=J_{ff}$.

It is important to notice that, as mentioned in the previous
section, the area is also (the square root of) the norm of its
associated \emph{selfdual}, or, equivalently, antiselfdual bivector:
\begin{equation}
A_f = 2 \sqrt{{}^+ \Sigma_f^{i}(t){}^+ \Sigma_f^{i}(t)} = 2 \sqrt{{}^-
\Sigma_f^{i}(t){}^- \Sigma_f^{i}(t)}. \label{areasd}
\end{equation}
These equalities are assured by the simplicity constraints on $\Sigma_f$.
Similarly,
\begin{equation}
J_{ff'} =  4{}^+ \Sigma_f^{i}(t){}^+ \Sigma^i_{f'}(t) =  4{}^- \Sigma_f^{i}(t){}^-
\Sigma^i_{f'}(t), \label{angle}
\end{equation}
Again, these
equalities follow from the simplicity constraints.

There exist a number of relations among the quantities $J_{ff'}$
within a single 4-simplex.\footnote{As a discussion of these
relations does not seem to appear elsewhere in the literature, we
have here included an appendix on them (\ref{app4s}).} Only 10 of
these quantities can be independent, because the geometry of a
4-simplex is determined by 10 numbers.  In particular, all angles
must be given functions $J_{ff'}(A_f)$ of the 10 areas (up to
degeneracies).   Explicit knowledge of the form of these functions
would be quite useful in quantum gravity.

Finally, consider now a triangle $f$.  Let $t_1, t_2, ... ,t_n$ be
the set of the tetrahedra in the link around $f$ and $v_{12},
v_{23}, ... ,v_{n1}$ be the corresponding set of simplices in this
link, where $t_2$ bounds $v_{12}$ and $v_{23}$ and so on cyclically.
In general, if we parallel transport  $e(t)$ across simplices around
a triangle, using $U_{tt'}$, we come back rotated, because of the
curvature at the triangle (the analog of a parallel transport around
the tip of a pyramid in 2d.)  In other words, we can always gauge
transform $e_{a}^I(t)$ to $\delta_{a}^I$ within a single cartesian
coordinate patch, but in general there is no cartesian coordinate
patch around a face.  We define
\begin{equation}
U_f(t_1) \equiv V_{t_1v_{12}}\ V_{v_{12}t_2} \ ...\ V_{t_n v_{n1}}V_{v_{n1}t_1}
\label{holonomieV}
\end{equation}
or, equivalently
\begin{equation}
U_f(t_1) \equiv U_{t_1 t_2}(v_{12}) \ ... \ U_{t_n t_1}(v_{n1}) ,
\label{holonomieU}
\end{equation}
the product of the rotation matrices obtained turning along the link
of the triangle $f$, beginning with the tetrahedron $t_1$. The
rotation matrix $U_f(t_1)$ then represents the $SO(4)$ curvature
associated to the triangle $f$, written in the $t_1$-frame.

\subsection{Dynamics}

The last step before attacking the quantization of the model is to 
write the discretized action. Take the $e(t)$ as independent 
variables so that $\Sigma_f(t)=e(t)\wedge e(t)$. In analogy with 
Regge calculus we define the action to be
\be S_{bulk}[e(t),U,V] = \frac{1}{2}\sum_{f}  \ Tr[B_f(t) U_{f}(t)]
+ \sum_v \sum_{f\subset v} Tr[\lambda_{vf} U_{tt'}(v) \; V_{t'v} \;
V_{vt}] . \label{actionF1}
\ee
where we recall that $B={}^*\Sigma$.
The first sum is over all faces in the interior of the
discretization and $U_f(t)$ is defined as in (\ref{holonomieU}),
where $t$ is any one of the tetrahedra in the link of $f$. We impose
a priori equation (\ref{issa}), which implies
\footnote{It follows that, for each $f$, there is only one
independent $B_f(t)$ in the link.  This is consistent with the
action (\ref{actionF1}), where only one $\Sigma_f(t)$ in each link
appears.  Note it follows that we are also imposing a priori
\begin{equation}
\Sigma_f(t) U_f(t) = U_f(t) \Sigma_f(t),
\end{equation}
so that not even this one independent $\Sigma_f(t)$ in the link is a
priori independent of the connection degrees of freedom.}
\begin{equation}
\Sigma_f(t) U_{tt'} = U_{tt'} \Sigma_f(t') .
\end{equation}
It follows that the first term in the action (\ref{actionF1}) is
independent of the choice of tetrahedron $t$ in the link of each
$f$.

The second term in (\ref{actionF1}) is a sum over all four simplices
and in each four simplex over all faces belonging to it
\footnote{This is equivalent to summing over the wedges defined by
Reisenberger in \cite{mike1}. Each wedge is identified by a pair
$(v,f)$.}. $\lambda_{vf}$ is a lagrange multiplier living in the
algebra of $SO(4)$ and varying with respect to it gives the
constraint (\ref{flat}) on the group variables.

This action is invariant under the gauge transformations \Ref{gaugetra} and
reduces to the action of GR in the limit in which the triangulation is fine.  
This can be seen as
as follows. In the limit in which curvatures are small, $U_f(t)=1+\frac{1}{2}F_{ab}dx^a\wedge dx^b$, where $dx^a\wedge dx^b$ is the plane normal to the triangle $f$. Hence the trace gives
\be
\frac{1}{2}Tr[B_f(t) U_{f}(t)]\sim \frac{1}{4}B^{IJ}_{ab} F^{KL}_{cd}\epsilon^{abcd}=
\frac{1}{4}\epsilon^{IJ}{}_{KL} e^K_a e^L_b F^{KL}_{cd}\epsilon^{abcd}=
e e^a_I e^b_J F^{IJ}_{ab} = \sqrt{g} R,
\label{actiontetrads}
\ee
which is the GR lagrangian density.

There is also a close relation of (\ref{actionF1}) to the Regge action.
To see this, let us first see how to extract the deficit angle around
each face in our framework.  Let $v_1^{\mu}, v_2^{\mu}$ denote two of
the edges of the triangle $f$.
%
%
%
Now, as the triangle $f$ is the axis of the parallel transport around $f$,
the \textit{tangent space} parallel transport map $U_f(t)^{\mu}{}_{\nu}$
preserves each of the vectors $v_1^{\mu}, v_2^{\mu}$:
\begin{equation}
U_f(t)^{\mu}{}_{\nu}v_i^{\nu} = v_i^{\mu}, \; i \in \{1,2\}.
\end{equation}
Contracting both sides with $e^I_{\mu}$, and inserting the resolution
of unity $e^{\nu}_J(t)e^J_{\rho}(t) = \delta^{\nu}_{\rho}$ we get
\begin{equation}
U_f(t)^I{}_J v_i^J = v_i^I, \; i \in \{1,2\}
\end{equation}
where $v_i^I:= e^I_{\mu}v_i^{\mu}$.  By linearity, $U_f(t)^I{}_J$ therefore
acts as identity on the subspace $V:= {\rm span}\{v_1^I, v_2^I\}$, forcing
$U_f(t)^I{}_J$ to belong to the $SO(2)$ subgroup fixing $V$.
Therefore $U_f(t)^I{}_J$ is described by a single angle: this is the
deficit angle around $f$.
To make this explicit, as well as to cast (\ref{actionF1}) directly
into Regge-like form, let us introduce an orthonormal basis
$\xi_1^I, \xi_2^I$ of the orthogonal complement of $V$.
The rotation $U_f(t)^I{}_J$ is a rotation in the $\xi_1$-$\xi_2$ plane.
Therefore
\begin{equation}
U_f(t)_{21} := \left(\xi_2\right)_I U_f(t)^I{}_J \xi_1^J = \sin \theta_f
\end{equation}
where $\theta_f$ is the angle by which $\xi_1$ is rotated by $U_f(t)$ ---
the deficit angle.
We have also
\begin{equation}
U_f(t)_{12} := \left(\xi_1\right)_I U_f(t)^I{}_J \xi_2^J = -\sin \theta_f .
\end{equation}
Furthermore,
\begin{equation}
B_f(t)^{IJ} \propto \xi_1^{[I}\xi_2^{J]} .
\end{equation}
The norm of $B_f(t)^{IJ}$ is just the area (\ref{area}), which fixes
the constant of proportionality here. We have
\begin{equation}
B_f(t)^{IJ} =  A_f \xi_1^{[I}\xi_2^{J]} .
\end{equation}
Thus
\begin{equation}
B_f(t)^{IJ} U_f(t)_{IJ} = A_f \xi_1^{[I}\xi_2^{J]} U_f(t)_{IJ}
= 2A_f \sin \theta_f .
\end{equation}
The bulk term in (\ref{actionF1}) can therefore be written
\begin{equation}
\sum_f A_f \sin \theta_f.
\end{equation}
To lowest order in $\theta_f$, this is precisely the bulk Regge action.

\subsection{Classical equations of motion}

From the form of the action (\ref{actiontetrads}), one can see 
that variation w.r.t. the tetrads will give the discrete analog of
the Einstein equations. Variation with respect to the connection is
more subtle. In order to proceed, let us write the action on shell
w.r.t. the constraints on the group variables so that it reads
$S[e,V]=\half \sum_f Tr[B_f(t)U_f(t)]$ where $U_f(t)$ is written as
in (\ref{holonomieV}). The action, as we saw, is independent of the
choice of the base tetrahedron in each face. Let us consider the
variation with respect to $V_{tv}$. This variable appears in four
terms in the sum, corresponding to the four faces of the tetrahedron
$t$. Let us choose in addition, for the sake of simplicity, the
tetrahedron $t$ to be the base of these four faces. The variation
($\delta V_{tv} = \xi V_{tv}$)
\footnote{The variation so defined is that determined by the right
invariant vector field associated to $\xi$.}
gives
\be \delta S=\half \sum_{f_i,i=1..4}Tr[B_{f_i}(t)\xi U_{f_i}(t)] \ee
where $\xi$ is an arbitrary infinitesimal element in the algebra of
$SO(4)$. Stationarity w.r.t. these variations implies that the
antisymmetric part of $\left[\sum_{f_i} U_{f_i}B_{f_i}\right]$, seen
as a four dimensional matrix, is zero. Explicitly,
\be
\sum_{f_i}U_{f_i}B_{f_i}+B_{f_i}U_{f_i}^{-1} =0.
\ee

Now, $U_f(t)$ can be expanded as $U_f \sim 1+u_f$ where $u_f$ is of
second order in the lattice spacing \footnote{Lattice spacing is
defined with respect to a particular continuum limit.  To be
specific, suppose one has a metric $g$ which one wishes to
approximate. Then, for each real number $\lambda$, one can construct
a corresponding Regge geometry $\Delta_\lambda$ approximating $g$,
such that the typical lattice spacing is $\lambda$, as measured
w.r.t. $g$.  As $\lambda \rightarrow 0$, $\Delta_\lambda$ approaches
$g$, and one can consider expansions in $\lambda$.}, so that to first 
order we get simply,
\be
\sum_{f_i}B_{f_i}=0
\ee
which is the closure constraint. The analogous continuum equation is 
the Gauss constraint, given by:

\be
DB=dB+[A,B]=0\label{gauss}
\ee
where $A$ is the spin connection. Remembering that the tetrahedron 
is flat, one can choose the connection to be identically zero. Integration 
over the region defined by $t$ gives the closure. In order to make this 
relation more clear, let us consider now the case where $t$ is not the 
base tetrahedron for all faces. Consider then, for the face $f_1$ the 
base tetrahedron to be $t_1$, where $t,t_1,...,t_n$ is the link around 
$f_1$. Variation w.r.t. $V_{tv}$ gives:

\be
\left[U_{tt_1}B_{f_1}(t_1)U_{tt_1}^{-1}U_{f_1}(t)+...+B_{f_4}(t)U_{f_4}\right]_{anti}=0
\ee
where $U_{tt_1}=V_{tv_1}V_{v_1t_1}$. To first order it reads:

\be
B_{f_1}(t_1) + [u_{tt_1},B_{f_1}(t_1)] + ... +B_{f_4}(t)=0
\ee
which can be seen directly as the integration of the equation 
(\ref{gauss}) over the region defined by $t$, where the spin 
connection is a distribution concentrated at the face $f_1$ 
in the interior of $t$.

\subsection{Topological term}\label{topos}

Recall that in the LQG approach the action that is quantized is the
Holst action \cite{holst}, obtained adding to the
action a topological term that doesn't change the equations of motion
\be
S_{Holst}=\int\;*(e\wedge e)\wedge F +\frac{1}{\gamma}\int\;
(e\wedge e)\wedge F.
\label{holstac}
\ee
where $\gamma$ is called the Immirzi parameter. The introduction
of a topological term is \textit{required} in order to have a
theory of connections on the boundary: without it, as shown by
Ashtekar, the connection variable does not survive the Legendre
transform \cite{abhay1991}. It is easy to add a similar
term  in the discrete theory, having no effect on the equations of motion
in the continuous limit.  This is
\be S_{top}[e(t),U,V] = \frac{1}{\gamma}\sum_{f}  \ Tr[\Sigma_f(t) U_{f}(t)].
\label{topo}
\ee
The reason this has no effects on the equations of motion is interesting.
In a Regge geometry, the curvature associated to the face $f$ is given by the
rotation $U_{f}(t)$. This rotation has the property of leaving the face $f$
itself invariant.  Hence, it is a rotation that is generated by the dual of the face
bivector $\Sigma_f(t)$ itself. That is, it has the form $U_{f}(t)\sim exp\{\theta B_f(t)\}$.
In the weak field limit, this gives  $U_{f}(t)\sim 1+\theta B_f(t)$, and therefore
$Tr[\Sigma_f(t) U_{f}(t)]\sim  \theta Tr[\Sigma_f(t) B_{f}(t)=
\theta \Sigma_f(t) {}^*\Sigma_{f}(t)]$ which vanishes because of
the simplicity constraint (because $\Sigma$ is simple).

This term will play a role in the quantization.

\subsection{Boundary terms}

In ordinary Regge calculus boundary terms must be added to 
the action so that the equations of motion are the same when 
we vary the action with some boundary variables fixed. The 
other condition for boundary terms is that they add correctly. 
Consider a region of spacetime which is the union of two 
disjoint regions separated by a boundary, $\rd{S}=\rd{S}'\cup \rd{S}''$. 
The additivity condition is then that $S[\rd{S}]=S[\rd{S}']+S[\rd{S}'']$ 
\cite{hartlesorkin}.

\begin{figure}
\begin{center}
  \includegraphics[height=4.5cm]{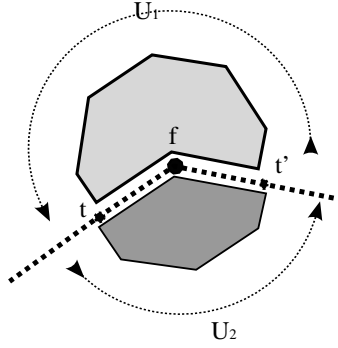}
\end{center}
\caption{The face $f$ and the two tetrahedra $t$ and $t'$ are on the boundary of the triangulation, represented by the heavy dashed line. The link of the face is divided  by the boundary into two parts.}
\label{reggebound}
\end{figure}

Consider for simplicity the part of the action (\ref{actionF1}) referring to a single face $S_f=Tr[B_f(t)U_f(t)]$. Furthermore, choose two tetrahedra $t$ and $t'$ appearing in the link around this face and break the link in two so that the two tetrahedra belong now to the boundary (cf. Figure \ref{reggebound}). The action can be split to first order in the algebra as
\be
S_f = \half Tr[B_f(t)U_{tt'}^{1} R_f] + \half Tr[B_f(t')U_{t't}^{2} R_f^{-1}]
\ee
where in the second term of the sum we have replaced $B_f(t)$ by $B_f(t')$ which, because of (\ref{issa}), introduces terms of the second order in the lattice spacing. $U_{tt'}^{(1,2)}$ are just the boundary connection variables. Here $R_f$ is a fixed $\SO(4)$ element inserted to make the terms $\SO(4)$ covariant.
Thus, the boundary action can be written in general as
\be
S_{\partial\Delta}=\half \sum_{f\subset \partial\Delta} Tr[B_f(t)U_{tt'}R_f]
\label{genboundaction}
\ee
with $U_{tt'}$ given by
\be
U_{tt'}= U_{tt_1}(v_1) \ ... \ U_{t_n t'}(v_n) .
\ee
This gives additivity to first order in the algebra.

Let us remark on the specific case when the $B$'s are fixed on the boundary.
In this case, in the continuum Plebanski theory, the boundary term is just $\int B\wedge A$ where
$A$ is the connection \cite{danieleboundary}. If we choose $R_f \equiv 1$, the above discrete boundary
action becomes
\begin{equation}
S_{\partial\Delta}=\half \sum_{f\subset \partial\Delta} Tr[B_f(t)U_{tt'}]
\label{Bboundaction}
\end{equation}
which reduces precisely to the continuum boundary action. 
Thus we choose $R_f \equiv 1$ when fixing
the $B$'s on the boundary.  The boundary action is not 
gauge-invariant,
as it is not in the continuum theory.\footnote{
One can also fix the $U$'s on the boundary.  In this case, 
there is no boundary term in the continuum
theory.  This can effectively be achieved in the discrete 
theory by setting $R_f = U_{tt'}^{-1}$
\textit{after} varying the action to obtain the equations of 
motion (see appendix \ref{appsymp}).}

Let us now write the action for a single four simplex as it 
will be useful to us in the next section. All the faces are 
on boundary and so the action is just a sum of boundary terms. Explicitly:
\be
S_{v}= \half \sum_{f\subset v}\; Tr[B_f(t_A)\; U_{t_A t_B}(v)] 
+ Tr[\lambda_f \; U_{t_A t_B}(v)\; V_{t_B v}\; V_{v t_A}]
\label{action4s}
\ee

\subsection{Boundary variables}

Suppose the triangulation $\Delta$ has a boundary $\partial\Delta$.
This boundary is a 3d manifold, triangulated by tetrahedra separated
by triangles.   Notice that, unlike what happens in the bulk, each
boundary tetrahedron bounds just a single simplex of the
triangulation; and each boundary triangle bounds just two boundary
tetrahedra.

Let us identify the boundary variables, which reduce to the boundary
gravitational field in the continuum limit. One boundary variable is
simply $\Sigma_f(t)$, where $f$ is a boundary triangle and $t$ is a
boundary tetrahedron.  There are only two boundary tetrahedra around
$f$, one, $t$, at the start of the link as determined by the
orientation of $f$, and the other, $t'$, at the end. It will turn
out to be convenient to use the notation $\Sigma^R_f$ and $\Sigma^L_f$ for
$\Sigma_f(t)$ and $\Sigma_f(t')$.  If desired, one can also associate
variables with the reverse orientation of $f$: $\Sigma^R_{f^{-1}}:=
\Sigma^L_f$, $\Sigma^L_{f^{-1}}:=\Sigma^R_f$.

The other boundary variable is the group element $U_f = U_{tt'}$
giving the parallel transport across each triangle $f$ bounding $t$
and $t'$ (not to be confused with the holonomy (\ref{holonomieV})
defined above). Notice that \Ref{issa} implies
\be
U_{f}\Sigma^R_fU_{f}^{-1}=\Sigma^L_f.
\label{BU2}
\ee

Finally, of the constraints (\ref{closurev}, \ref{aa}, \ref{bb}, \ref{zucchero1},
\ref{flat}), only (\ref{closurev}, \ref{aa}, \ref{bb},  \ref{zucchero1}) act
separately at each tetrahedron, and thus impose direct restrictions
on boundary data. We call (\ref{closurev}, \ref{aa}, \ref{bb}, \ref{zucchero1}) the
kinematical constraints. The constraint (\ref{flat}) on the other
hand necessarily involves bulk variables; we thus call it the
dynamical constraint.

The complex dual to the boundary triangulation defines a 4-valent
graph $\Gamma$ with nodes $t$ (dual to the boundary tetrahedron $t$)
and links $f$ (dual to the boundary triangle $f$). It is convenient
to view the boundary variables as associated to the four valent
graph $\Gamma$ dual to the boundary triangulation: we have $SO(4)$
group elements associated to the links of the graph and two $\Sigma$
variables associated to the two orientations of each such link.
Accordingly, we change notation, and call $l$ (for links) the
oriented boundary triangles and $n$ (for nodes) the boundary
tetrahedra.

Since we are in a first order formalism, the space of these
variables $(\Sigma_l^L, \Sigma^R_l, U_l)$ code the phase space of
discretized GR.   In fact, this space is precisely the same as the
phase space of a Yang-Mills $SO(4)$ lattice theory, and it can be
identified as the cotangent bundle of the configuration space
$C=SO(4)^L$, where $L$ is the number of links on $\Gamma$.

This cotangent bundle has a natural symplectic structure, which
defines  Poisson brackets.
\begin{eqnarray}
\label{pbUUflip}
\{ U_l ,U_{l'} \} &=& 0  \nonumber \\
\label{pbBLUflip}
\{ (\Sigma^L_l)^{IJ} ,U_{l'} \} &=& \delta_{ll'}\  U_l\ \tau^{IJ} \nonumber\\
\label{pbBRUflip}
\{ (\Sigma^R_l)^{IJ} ,U_{l'} \} &=& \delta_{ll'}\  \tau^{IJ}\ U_l\ \nonumber \\
\label{pbBBflip}
\{ (\Sigma^R_l)^{IJ} ,  (\Sigma^R_l)^{KL}  \} &=& \delta_{ll'}\
\lambda^{IJ\,KL}{}_{MN} (\Sigma_l^R)^{MN}
\label{sympstflip}
\end{eqnarray}
where $\tau^{IJ}$ and $ \lambda^{IJ\,KL}_{MN} $ are, respectively,
the generators and the structure constants of $SO(4)$.
Observe that the two quantities $\Sigma^R_l$ and $\Sigma^L_l$ act very
nicely as the right,  and respectively, left invariant vector fields
on the group. Equation \Ref{BU2} gives the correct transformation
law between the two.

However, it is important to notice a crucial detail at this point.  As pointed
out in \cite{Montesinos:2006qg}
and in \cite{baezbarrett}, because of its peculiar $SU(2)\times
SU(2)$ structure, the cotangent bundle over $SO(4)$ carries indeed
\emph{two} different natural symplectic structures, related to one another by
what Baez and Barrett call a \emph{flip}:  one is obtained from the other by
flipping the sign of the antiselfdual part.  That is, replacing $\Sigma$  by $B$
\begin{eqnarray}
\label{pbUU}
\{ U_l ,U_{l'} \} &=& 0  \nonumber \\
\label{pbBLU}
\{ (B^L_l)^{IJ} ,U_{l'} \} &=& \delta_{ll'}\  U_l\ \tau^{IJ}\nonumber \\
\label{pbBRU}
\{ (B^R_l)^{IJ} ,U_{l'} \} &=& \delta_{ll'}\  \tau^{IJ}\ U_l\ \nonumber \\
\label{pbBB}
\{ (B^R_l)^{IJ} ,  (B^R_l)^{KL}  \} &=& \delta_{ll'}\
\lambda^{IJ\,KL}{}_{MN} (B_l^R)^{MN}
\label{sympst}
\end{eqnarray}
Both structures are equivalent to the lattice Yang Mills theory Poisson brackets \cite{creutz},
the difference is only whether we identify the electric field with $B$ or with $\Sigma$.
We call (\ref{sympstflip}) the ``flipped" Poisson structure (at the risk
of some confusion, since Baez and Barrett call (\ref{sympst}) the ``flipped"
Poisson structure. Being flipped as everything else, is a relative notion...)

Which one is the correct Poisson structure to utilize? The classical equations of motion,
which is the part of the theory that is empirically supported,  do not determine the symplectic structure uniquely.  In the Appendix \ref{appsymp}
we study the direct construction of the symplectic structure from the action.  If we take
the action \Ref{actionF1} without the topological term \Ref{topo}, then we have
the unflipped  Poisson structure (\ref{sympst}). (This is why we call it ``unflipped".)
But in order to arrive
at the Ashtekar formalism and LQG, we know that the topological term is
needed. As shown in the appendix, the two symplectic structures written above
are recovered by taking $\gamma\gg 1$ and $\gamma \ll 1$, respectively.  We discard
the first choice that gives macroscopic discrete area eigenvalue and we choose the
second.\footnote{Intermediate choices may also be of interest, but they lead to
a more complicated quantum theory, involving also non-simple representations. See
Appendix  \ref{appsymp}.}

Thus, we choose the \emph{flipped} symplectic
structure  (\ref{sympstflip}) as a basis for the quantization of theory.
This choice leads to a nontrivial intertwiner space, while the  opposite choice leads
to the Barrett-Crane trivial intertwiner space, as we shall see below.

\subsection{Summary of the classical theory}

Summarizing, discretized GR can be defined by the action
\Ref{actionF1} with the appropriate boundary terms
(\ref{genboundaction}) now defined as a function of the variables
$[\Sigma_f(t),V_{vt},U_{tt'}(v),\lambda_{vf} ]$). That is
\be S_{bulk}[\Sigma,U,V,\lambda] = \frac{1}{2}\sum_{f}  \ Tr[B_f(t) U_{f}(t)]
+ \sum_v \sum_{f\subset v} Tr[\lambda_{vf} U_{tt'}(v) \; V_{t'v} \;
V_{vt}] . \label{actionF11}
\ee
Plus the diagonal and off-diagonal simplicity constraints
\begin{eqnarray}
&&C_{ff}:=\frac{1}{4} {}^*\Sigma_f(t)\cdot \Sigma_f(t)= 0,\label{simplicitya} \\
&&C_{ff'}:=\frac{1}{4} {}^*\Sigma_f(t)\cdot \Sigma_{f'}(t)= 0. \label{simplicityb}.
\end{eqnarray}
The second can be replaced by the (stronger) condition that there is an $n_I$ for each
$t$ such that
\be
n_I\  \Sigma_f^{IJ}(t)=0,
\label{eccola0}
\ee
equivalently, there is a gauge in which
\be
\Sigma_f^{0i}(t)=0.
\label{eccola}
\ee
The following two other constraints follow from the equations of motion
\begin{eqnarray}
&&\sum{}_{f\in t} \ \Sigma_f(t)= 0, \label{fermeture}
\end{eqnarray}
and
\be
U_{t_A t_B}(v)=V_{t_A v}\; V_{v t_B} .
\label{last}
\ee


\section{Quantization}

The quantization of the theory will proceed in three steps. First,
we write the Hilbert space and the operators that quantize the
boundary variables and their Poisson algebra.  Second, we impose the
kinematical constraints
(\ref{fermeture},\ref{simplicitya},\ref{simplicityb},\ref{eccola}).
The first two of these pose no problem.  The third one, namely the
nondiagonal simplicity constraints (\ref{simplicityb}) require a
careful discussion and some technical steps. As we shall see, we
cannot impose these constraints strongly; we impose them weakly, in
a sense defined below.  Finally, the dynamics is specified by
computing an amplitude for a state in the boundary Hilbert space
\cite{oeckl,carloproj}.  This amplitude is constructed by building
blocks, the elementary building block being a single four-simplex.

Remarkably, we will find that the physical Hilbert space that solves
the constraints is naturally isomorphic to the kinematical Hilbert
space of $\SO(3)$ LQG.  More precisely, it is isomorphic to the set
of states of $SO(3)$ LQG defined on the graph formed by the dual of
the boundary triangulation. Our key result will then be a transition
amplitude for a state in the Hilbert space of $SO(3)$ spin networks
defined on the graph corresponding to the (dual of the) boundary of
a four-simplex.

\subsection{Kinematical Hilbert space and operators}

The boundary phase space is the same as the one of an $SO(4)$ lattice
Yang-Mills theory. We quantize it in the same manner as
$SO(4)$ lattice Yang-Mills theory  \cite{creutz}.  That is, the natural quantization of
the symplectic structure of the boundary variables is defined on the
Hilbert space $\Hil_{\SO(4)}:=L_2[\SO(4)^L]$, formed by
wave functions $\psi(U_l)=\psi( g^+_l, g^-_l)$.  The $U_l$ variables
are quantized as diagonal operators. The variables $\Sigma^L_l$ and
$\Sigma^R_l$ are quantized as the left invariant ---respectively right
invariant--- vector fields on the group.  These define a representation
of the flipped classical Poisson algebra  (\ref{pbUUflip}).

A basis in this space is given by the states \be
     \psi_{j_l^+ j_l^-I_n^+ I_n^-}( g^+_l, g^-_l)= \langle g^+_l g^-_l | j_l^+, j_l^-,I_n^+, I_n^-\rangle=
\left(\bigotimes_l   D^{(j^+_l)}(g^+_l) \cdot \bigotimes_n
I^+_n\right)\ \left(\bigotimes_l   D^{(j^-_l)}(g^-_l) \cdot
\bigotimes_n I^-_n\right). \label{so4sn} \ee Here $D^{(j)}(g)$ the
are matrices of the $SU(2)$ representation $j$: two of these are
associated to each link; together, they form the $SO(4)$
representation matrix $(j^+,j^-)$, defined on the Hilbert space
$\rd{H}_{(j^+,j^-)}$.  At each node, $I_n^+ \in \Hil_{j^+_1} \otimes
\cdots \otimes \Hil_{j^+_4}$ and $I_n^- \in \Hil_{j^-_1} \otimes
\cdots \otimes \Hil_{j^-_4}$, so that $I_n^+ \otimes I_n^-$ defines
an element of the space \be
\mathcal{H}_{(j^+_1,j^-_1)...(j^+_4,j^-_4)}:=\left(\rd{H}_{(j^+_1,j^-_1)}\otimes
... \otimes\rd{H}_{(j^+_4,j^-_4)}\right) \ee associated to each node
$n$ with fixed adjacent representations $(j^+_1,j^-_1)...(j^+_4,j^
-_4)$. The contraction pattern of the indices between the
representation matrices and the tensors  $(I^+, I^-)$ is defined by
$\Gamma$.

Next, we consider the constraints.

Strictly speaking, the closure constraint \Ref{fermeture} does not
need to be imposed at this stage, since it is not an independent
constraint, but it is implemented by the dynamics, as shown in the
last section.  But, anticipating, let us see what is its effect. It
reduces $L_2[SO(4)^L]$ to the space $L_2[SO(4)^L/SO(4)^N]$ formed by
the states invariant under
$\psi(U_l)=\psi(V_{s(l)}U_lV_{t(l)}^{-1})$.   That is, it constrains
the tensors $(I^+, I^-)$ to be $SO(4)$ intertwiners $(i^+,i^-)$.
That is, it constrains $I^+ \otimes I^-$ to belong to the subspace
$\mathcal{K}^{SO(4)}_{t}:= \mathcal{K}^{SO(4)}_{(j_1^+,j_1^-) \dots
(j_4^+,j_4^-)}:={\rm Inv}\left(\rd{H}_{(j^+_1,j^-_1)}\otimes ...
\otimes\rd{H}_{(j^+_4,j^ -_4)}\right)$, which is the space of the
$SO(4)$ intertwiners. The state space obtained by imposing the
closure constraint is then precisely the Hilbert space of an $SO(4)$
lattice Yang-Mills theory.

The diagonal simplicity constraint \Ref{simplicitya} restricts to the spin network states where the representation associated to the links is simple. That is, it imposes $j_l^+= j_l^-\equiv j_l$.

Let us now come to the off-diagonal simplicity constraints \Ref{simplicityb},
which are of central interest to us.
After \Ref{fermeture} and \Ref{simplicitya} are satisfied,
only two of the three off-diagonal simplicity constraints
acting on each tetrahedron are independent.
These constraints form a second class system.
Imposing them strongly restricts the space of intertwiners to one unique
solution given by the Barrett-Crane vertex.\footnote{
The commutator of two of these constraints is called
chirality in \cite{BC1}.   In \cite{BC1}, the chirality constraint
is imposed strongly on the states as well, with the result of
selecting the non-degenerate geometries corresponding to
the sector \Ref{sol1}.  Notice, however, that the system formed by the
the chirality and the simplicity constraints
is not first class either, as the chirality, in turn, does not
commute with the simplicity constraints.}

In order to illustrate the problems that follow from imposing second
class constraints strongly, and a possible solutions to this
problem, consider a simple system that describes a single particle,
but using twice as many variables as needed. The phase space is the
doubled phase space for one particle, i.e., $T^* \mathbb{R}\times
T^*\mathbb{R} \ni \left((q_1 ,p_1),(q_2 ,p_2)\right)$, and the
symplectic structure is the one given by the commutator
$\left\{q_a,p_b\right\}=\delta_{ab}$. We set the constraints to be
\begin{eqnarray}
q_1-q_2=0, \nonumber \\
p_1-p_2=0.
\end{eqnarray}
By defining the variables $q_{\pm}=(q_1 \pm q_2)/2$ and
$p_{\pm}=(p_1 \pm p_2)/2$, the constraints read: $q_{-}=p_{-}=0$.
They are clearly second class since $\left\{q_{-},p_{-}\right\}=1$.
Suppose we quantize this system on the Schr\"odinger Hilbert space
$L_2[R^2]$ formed by wave functions of the form $\psi(q_+,q_-)$. If
we impose the two constraints strongly we obtain the set of two
equations
\begin{eqnarray}
q_-\  \psi(q_+,q_-)=0, \nonumber \\
i\hbar \frac{\partial}{\partial q_-}\  \psi(q_+,q_-) = 0.
\end{eqnarray}
which has no solutions.  We have lost entirely the system.

There are several ways of dealing with second class systems.   
One possibility, which is employed for instance in the 
Gupta-Bleuler formalism for electromagnetism and in 
string theory, can be illustrated as follows in the context 
of the simple model above  (see for instance the appendix 
of \cite{tate}).  Define the creation and annihilation 
operators  ${a}_-^\dagger=(p_{-}+iq_{-})/{\sqrt{2}}$ 
and $a_-=(p_{-}-iq_{-})/{\sqrt{2}}$. The constraints 
now read $a_-=a_-^\dagger=0$. Impose only one 
of these strongly:  $a_- |\psi\rangle=0$ and call the 
space of states solving this $\mathcal{H}_{ph}$.  
Notice that the other one holds weakly, in the sense  that
\begin{eqnarray}
\langle \phi| {a}_-^\dagger |\psi\rangle=0\hspace{3em} \forall\ \phi, \psi \in \mathcal{H}_{ph}.
\end{eqnarray}
That is, ${a}_-^\dagger$ maps the physical Hilbert 
space  $\mathcal{H}_{ph}$ into a subspace orthogonal 
to  $\mathcal{H}_{ph}$.  Similarly, in the Gupta-Bleuler 
formalism the Lorentz condition (which forms a second 
class system with the Gauss constraint) holds in the form
\begin{eqnarray}
\langle \phi| \partial^\mu A_\mu |\psi\rangle=0\hspace{3em} 
\forall\ \phi, \psi \in \mathcal{H}_{ph}.
\end{eqnarray}

A general strategy to deal with second class constraints is
therefore to search for a decomposition of the Hilbert space of the
theory $\mathcal{H}_{kin}=\mathcal{H}_{phys}\oplus\mathcal{H}_{sp}$
(sp. for spurious) such that the constraints map
$\mathcal{H}_{phys}\rightarrow\mathcal{H}_{sp}$. We then say that
the constraints vanish weakly on $\mathcal{H}_{phys}$. This is the
strategy we employ below for the off-diagonal simplicity constraints
$C_{ll'}$.  Since the decomposition may not be unique, we will have
to select the one which is best physically motivated.  We now define
this space.

\subsection{The physical intertwiner state space ${\rd K}_{ph}$}
\label{intspace_sect}

Consider the state space obtained by imposing the diagonal
simplicity constraint, namely by taking $j_l^+=j_l^-:=j_l$, but not
the closure constraint yet.  Let us restrict attention to a single
tetrahedron $t$.  The constraints  \Ref{simplicityb}  at $t$ act on
the space associated to $t$, which is
$\mathcal{H}_{(j_1,j_1)...(j_4,j_4)}$. The closure constraint will
then restrict this space to the $SO(4)$ intertwiner space
$\mathcal{K}^{SO(4)}_{t}$.   We search for a subspace ${\rd K}_{ph}$
of
 $\mathcal{K}^{SO(4)}_{t}$ where the nondiagonal constraints vanish weakly.

First, note $\mathcal{K}^{SO(4)}_{t} =
{\rm Inv}(\mathcal{H}_{(j_1,j_1)} \otimes ...\mathcal{H}_{(j_4,j_4)})$
is a subspace of the larger space
\begin{equation}
\mathcal{H}_{(j_1,j_1)} \otimes ...\mathcal{H}_{(j_4,j_4)}
= \mathcal{H}_{j_1 \otimes j_1} \otimes ...\mathcal{H}_{j_4 \otimes j_4}
\end{equation}
which can be thought of as a tensor product of carrying spaces of
$SO(4)$ representations or $SU(2)$ representations, as desired.
The Clebsch-Gordan decomposition for the first factor on the right-hand
side above gives
\begin{equation}
\mathcal{H}_{j_1 \otimes j_1} = \mathcal{H}_{j_1} \otimes \mathcal{H}_{j_1} =
\mathcal{H}_0 \oplus \mathcal{H}_1 \oplus \dots \oplus \mathcal{H}_{2j_1}
\end{equation}
and similarly for the other factors.  By selecting the highest spin
term in each factor, we obtain a subspace
\begin{equation}
\mathcal{H}_{2j_1} \otimes \dots \otimes \mathcal{H}_{2j_4} .
\end{equation}
Orthogonal projection of this subspace into $\mathcal{K}^{SO(4)}_t$
then gives us the desired $\rd{K}_{ph} \subset
\mathcal{K}^{SO(4)}_t$. This $\rd{K}_{ph}$ is the intertwiner space
that we want to consider as a solution of the constraints. The total
physical boundary space  ${\rd H}_{ph}$ of the theory is then
obtained as the span of spin-networks in $L_2[SO(4)^L/SO(4)^N]$ with
simple representations on edges and with intertwiners in the spaces
${\rd K}_{ph}$ at each node. Notice that the elements in ${\rd K}_{ph}$ 
are not necessarily simple in their
internal representation, in any basis. Let us study the properties of the 
space ${\rd K}_{ph}$, and
the reasons of its interest.
\begin{enumerate}
\item[(i)]
First, it is easy to see that the off-diagonal simplicity
constraints (\ref{simplicityb}) vanish weakly $\rd{K}_{ph}$, in the
sense stated above.   This follows from the following consideration.
A generic element of  ${\rd K}_{t}^{SO(4)}$  can be expanded as \be
\ket{\psi}=\sum_{i^+ i^-}\; c_{i^+ i^-} \ket{i^+ , i^-} \ee where
$i^\pm$ defines a basis in the $SU(2)$ intertwiner space. The
off-diagonal simplicity constraints are odd under the exchange of
$i^+$ and $i^-$, namely under exchange of self dual and antiselfdual
sectors. But the states in $\rd{K}_{ph}$ are symmetric in  $i^+$ and
$i^-$.  Hence $\langle \phi |C_{ll'}|\psi \rangle=0, \ \  \forall
\phi,\psi\in\rd{K}_{ph} $, that is,  $\rd{K}_{ph}$ can be considered
as one possible solution of the constraint equations.

\item[(ii)]
Second, let us motivate the choice of this solution. Recall that the
off-diagonal simplicity constraints can be expressed as the
requirement that there is a direction $n_I$ such that \Ref{eccola}
holds.  But this is precisely the ``classical" limit of the
condition satisfied by the spin-$2j$ representation, as observed at
the end of Section \ref{selfdual}. That is, promoting \Ref{eccola}
to the quantum theory gives the requirement that there there is a
gauge in which \be J^{0i}= 0. \ee which is \Ref{zumzum}, namely the
condition satisfied by the spin-$2j$ component of the representation
$(j,j)$. Equivalently, \Ref{eccola} implies
 \be 
 2C_{4} = \half
J_l^{IJ} J_{l\,IJ}= \half J_l^{ij} J_{l\,ij}= C_{3} \label{cas} 
\ee
where $C_4$ is the $SO(4)$ scalar Casimir in the representation
${\rd H}_{j_l,j_l}$ and $C_3$ is the Casimir of the $SO(3)$ subgroup
that leaves $n^I$ invariant, in the same representation. As it is,
this relation has in general no solution in ${\rd H}_{j_l,j_l}$, but
if we order it in slightly different manner, as in  \Ref{qcas1},
(reinstating $\hbar \ne 0$ for clarity)
\begin{equation}
C=\sqrt{C_3+{\hbar^2}/{4}}-
\sqrt{2C_4+\hbar^2}+{\hbar}/{2}=0
\label{qcas}
\end{equation}
then there is always a solution, which is given by the $H_{2j}$
subspace of $H_{j,j}$.   Therefore the off-diagonal simplicity
constraints pick up precisely the space we have defined. In other
words, this space satisfies a quantum constraint, which in the
classical limits becomes the classical constraint (\ref{cas}).

Notice that if we hadn't chosen the flipped symplectic structure,
then we would have obtained $J^{ij}=0$, instead of $J^{0i}=0$, that
is, the vanishing of the $SO(3)$ Casimir.  Following the same
procedure as above, we would have defined the space
\begin{equation}
\rd{K}^{(0)}_{ph}= {\rm Inv}\left(\rd{H}_{0}\otimes ... \otimes\rd{H}_{0}\right)\subset \mathcal{H}_{j_1...j_4}
\end{equation}
and this is precisely the one dimensional Barrett-Crane intertwiner space. That it,
it is the choice of the flipped symplectic structure that allows the selection of
a nontrivial intertwiner space.

\item[(iii)]
Third, we have the remarkable result that ${\rd K}_{ph}$ is
isomorphic to the $SO(3)$ intertwiner space, and therefore the
constrained boundary space ${\rd H}_{ph}$ is precisely the $SO(3)$
LQG state space  ${\rd H}_{SO(3)}$  associated to the graph which is
dual to the boundary of the triangulation, namely the space of the
$SO(3)$ spin networks on this graph.  We exhibit this isomorphism in
a way that simultaneously shows a new way of viewing ${\rd H}_{ph}$.
We will first construct a projection $\pi: {\rd H}_{SO(4)}\to {\rd
H}_{SO(3)}$. The hermitian conjugate map will then be an embedding
$f: {\rd H}_{SO(3)} \rightarrow {\rd H}_{SO(4)}$. The image of this
embedding will be none other than ${\rd H}_{ph}$.

The map $\pi:{\rd H}_{SO(4)}\to {\rd H}_{SO(3)}$ is simple to
construct. Choose an $SO(3)$ subgroup $H$ of $SO(4)$. As explained
in Section \ref{selfdual}, this choice is equivalent to the choice
of an isomorphism between the two $SU(2)$ factors of $SO(4)$. A
(simple) $SO(4)$ representation $\rd{H}_{(j,j)}$ is in general a
\emph{reducible} representation of $H$: since $H$ acts on each
duality component independently, it transforms as
$\rd{H}_{(j,j)}=\rd{H}_{j}\otimes\rd{H}_{j}$, where $\rd{H}_j$ is
the standard spin-$j$ representation of $SU(2)$. This decomposes
into irreducibles as
\begin{equation}
\rd{H}_{(j,j)}=\rd{H}_{0}\oplus \rd{H}_{1}\oplus ...\oplus \rd{H}_{2j},
\end{equation}
Thus, begin with an $SO(4)$ spin-network state $\Psi$, and consider
the restriction $\Psi |_{\overset{l}{\times}H}$ to the subgroup $H$
on each edge.  As $\Psi$ is $SO(4)$ invariant, it is in particular
$H$-invariant, and thus the restriction $\Psi
|_{\overset{l}{\times}H}$ is a sum of $SO(3)$ spin-networks.
Furthermore, because $\Psi$ is $SO(4)$ invariant, and because all
possible $SO(3)$ subgroups $H$ are related to each other by
conjugation, this sum of spin-networks is independent of the choice
of $SO(3)$ subgroup $H$. The above considerations tell us that this
sum is just that given by the above Clebsch-Gordan decomposition for
each edge. The result of the projection $\pi \Psi$ is then defined
to be the spin-network in this sum corresponding to \textit{the
highest weight spin on each edge.} Thus $\pi$ so-defined maps simple
$SO(4)$ spin-network states to $SO(3)$ spin-network states. The spin
$(j,j)$ on each edge of the $SO(4)$ spin-network is mapped to spin
$2j$ in the $SO(3)$ spin-network.  Each $SO(4)$ intertwiner $I_v \in
{\rm Inv}_{SO(4)} ({\rd H}_{(j_1,j_1)}\otimes ...\otimes {\rd
H}_{(j_4,j_4)})$ is mapped to an $SO(3)$ intertwiner simply via
orthogonal projection onto the subspace ${\rm Inv}_{SU(2)} ({\rd
H}_{(2j_1)}\otimes ...\otimes {\rd H}_{(2j_4)}) \subset ({\rd
H}_{(2j_1)}\otimes ...\otimes {\rd H}_{(2j_4)})$.

Let us now describe the conjugate embedding $f:{\rd H}_{SO(3)}\to
{\rd H}_{ph}$. This is defined as the hermitian conjugate of $\pi$,
using the fact that $\pi$ is a linear map between Hilbert spaces.
Let us describe it in detail.

Let us first describe the embedding $f$ restricted to a single
intertwiner space, namely the map (that we also call) $f$ from the
space ${\rd K}_{SO(3)}$ of the $SO(3)$ intertwiners to ${\rd
K}_{ph}$. Consider an intertwiner $i\in {\rd K}_{SO(3)}$, between
the four representations $(2j_1... 2j_4)$. Contract it with four
trivalent intertwiners between the representations $(2j_a, j_a,
j_a)$, the edge with spin $2j_1$ being contracted with the $2j_1$
edge of the corresponding trivalent intertwiner, etc.  This gives us
a tensor $e(i)$ in $({\rd H}_{(j_1,j_1)}\otimes ...\otimes {\rd
H}_{(j_4,j_4)})$. $e(i)$ is not an $SO(4)$ intertwiner, because it
is not $SO(4)$ invariant, but we obtain an $SO(4)$ intertwiner by
projecting it in the invariant part of $({\rd H}_{(j_1,j_1)}\otimes
...\otimes {\rd H}_{(j_4,j_4)}$). Since $SO(4)$ is compact, this
projection can be implemented by acting with a group element $U$ in
each representation, and integrating over $SO(4)$.
\be
f(i):=\int_{SO(4)}\; dV \; \left(\bigotimes_l\;
D^{(\lambda_l)}(V)\right)\; \cdot e(i) \label{emb2} \ee
The $SO(4)$ action can be factorized into two $SU(2)$ group
elements, one acting on the self dual, and other on the
anti-selfdual representations. One of the two factors can be
eliminated by virtue of the $SU(2)$ invariance of the trivalent
intertwiners and $i$.  What remains is an $SU(2)$ integration over
just one of the representations. Using the well known relation
\begin{equation}
\int_{SU(2)}  dg  D^{a_1b_1}(g) ...  D^{a_4b_4}(g)= \sum_i i^{a_1...a_4}i^{b_1...b_4}
\end{equation}
it is easy to see that we have
\begin{equation}
f|i\rangle = \sum_{i^+i^-} f_{i^+i^-}^i |i^+i^-\rangle
\end{equation}
where the coefficients $f_{i^+i^-}^i$ are given by the evaluation of the spin network
\begin{center}
  \includegraphics[height=3.5cm]{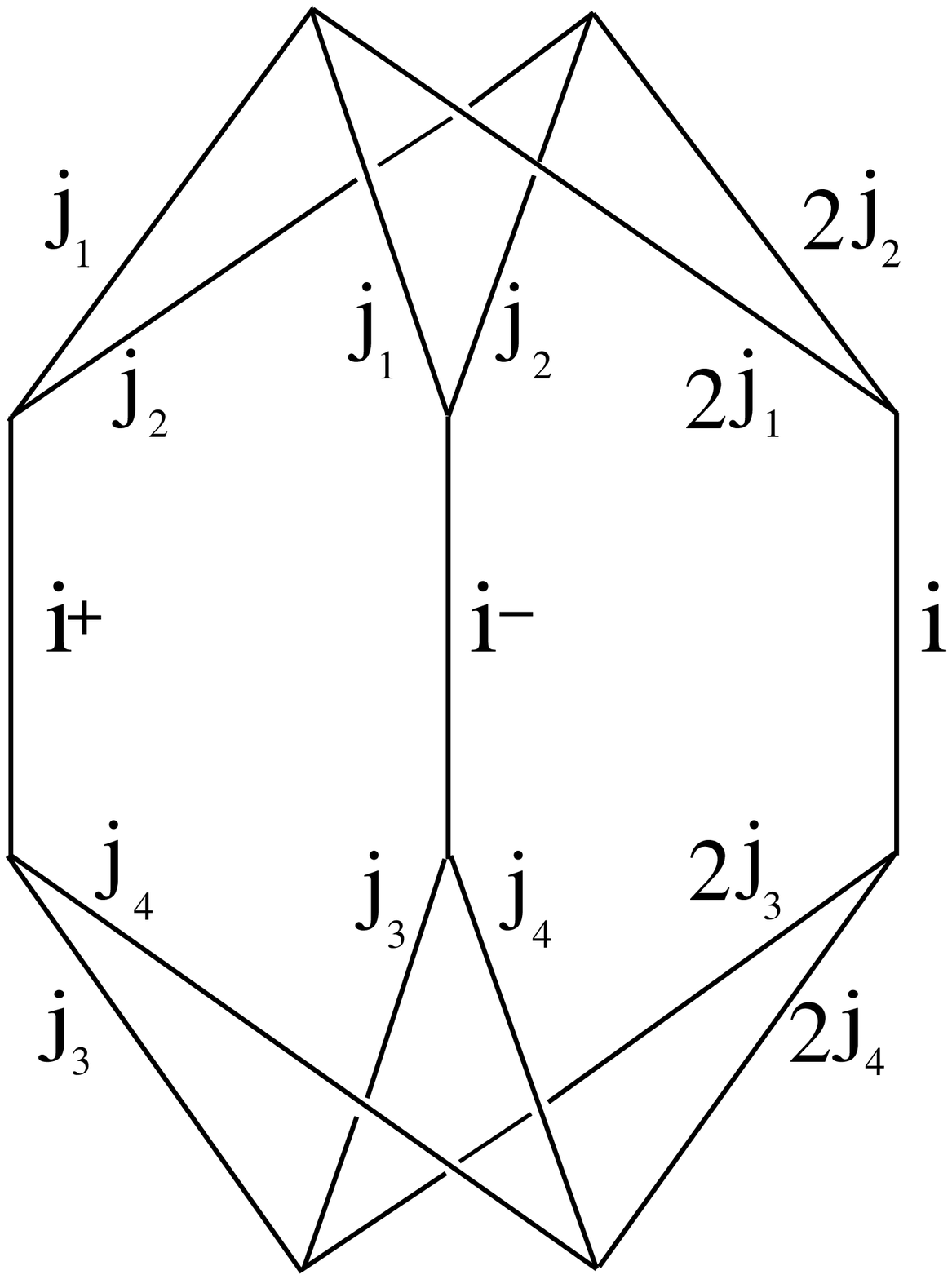}.
\end{center}

If we piece these maps at each node, we obtain the map $f:{\rd
H}_{SO(3)}\to {\rd H}_{ph}$ of the entire LQG space into the state
space of the present theory. In the spin network basis we obtain \be
f\; : |j_l,i_n\rangle  \longmapsto \sum_{i_n^+,i_n^-}
f_{i_n^+,i_n^-}^{i_n} |{j_l}/{2},{j_l}/{2},i_n^+,i_n^-\rangle.
\label{fi} \ee Equivalently, writing explicitly the states as
functions on the groups,
\begin{eqnarray}
f\; :\; \left(\bigotimes_l D^{(j_l)}(g_l)\right)\cdot\left(\bigotimes_n i_n\right) &\longmapsto&
\nonumber\\  &&
\hspace{-12em} \int_{SO(4)^N} \; \prod_n\; dV_n\; \left(\bigotimes_l D^{({\scriptstyle\frac{j_l}{2}},{\scriptstyle\frac{j_l}{2}})}\left(V_{s(l)}(g_l^+,g_l^-)V_{t(l)}^{-1}\right)\right)\cdot\left(\bigotimes_n e(i)_n\right)
\label{state_embed}
\end{eqnarray}
where indices have been omitted and $s(l),t(l)$ stand resp. for source and target of the link $l$. This completes the definition of the projection and
the corresponding embedding of the Hilbert space of LQG into the
boundary Hilbert space of the model.

\item[(iv)]
Let us illustrate more in detail the construction in (iii) using the
standard spinor notation  \cite{penroserindler}.  The vectors in
${\rd H}_j$ can be represented as totally symmetric tensors with
$2j$ spinor indices $(A_1...A_{2j})\equiv \rd{A}$, where each index
$A=0,1$ is in the fundamental representation of $SU(2)$. An element
in $\mathcal{H}_{(j_1,j_1) ...(j_4,j_4)}$ has therefore the form \be
I^{(A_1...A_{2j_1})(A'_1...A'_{2j_1})...(D_1...D_{2j_4})(D'_1...D'_{2j_4})}=:I^{\rd{A}\rd{A}'
...\rd{D}\rd{D}'} \ee here primed and unprimed indices are
symmetrized independently; they live in the self dual and antiself
dual components of the representation. Round brackets stand for
symmetrization. By choosing to no longer distinguish between primed
and unprimed indices, this $SO(4)$ intertwiner $I$ between the
simple representations $(j_1,j_1)...(j_4,j_4)$ becomes a tensor
among the $\SU(2)$ representations $j_1 \otimes j_1, j_2 \otimes
j_2, j_3 \otimes j_3, j_4 \otimes j_4$. Because of the
$\SO(4)$-invariance of $I$, the resulting $\SU(2)$ tensor does not
depend on the way primed and unprimed indices are identified.

Let us first construct the projection $\pi$, and corresponding
embedding $f$, for individual intertwiner spaces. The projection
$\pi : \mathcal{K}_{SO(4)}\rightarrow
\rd{K}_{SO(3)}$ is simply given by symmetrizing over the spinor
indices associated with each link
\be
\pi\; :\; I^{\rd{A}\rd{A}'...\rd{D}\rd{D}'} \longmapsto I^{(\rd{A}\rd{A}')...(\rd{D}\rd{D}')} =: i^{a...d}
\label{proj}
\ee
where the index $a$ is short for $(\rd{A}\rd{A'}):= A_1 \dots
A_{2j_1} A'_1\dots A'_{2j_1}$, and similarly for $b$,$c$, and $d$.
This symmetrization projects $I$ to an $\SU(2)$ intertwiner among
the representations $2j_1, 2j_2, 2j_3, 2j_4$, thereby selecting the
highest weight irreducible representation in the decomposition $j
\otimes j = 0 \oplus \cdots \oplus 2j$ on each edge.
Notice that $2j \in \Z$ so that the projected intertwiner transforms
under $SO(3)$ transformations.

Next, we write, for $j_1, \dots j_4 \in \Z$, the corresponding
embedding $f : \rd{K}_{SO(3)}\rightarrow\mathcal{K}_{SO(4)}$.
First, the embedding $e$ is trivially obtained by reading the
indices $\rd{A}\rd{A}'$ as living in self-dual and anti-self-dual
representations, respectively:
\be e\; :\; i^{a...d} \longmapsto
e(i)^{\rd{A}\rd{A}'...\rd{D}\rd{D}'}
:=i^{\rd{A}\rd{A}'...\rd{D}\rd{D}'} \label{emb1} \ee This is not yet
a $SO(4)$ intertwiner and in order to recover an element of the
invariant subspace we have to ``group average'' as in equation (\ref{emb2}).

Let us now consider the projection, and corresponding embedding, for individual
spins. Consider the state space for a single link -- that is, the space of square
integrable functions over one single copy of $SO(4)$. This space is spanned by 
the representation matrices 
$D^{(j^+,j^-)}\left(U_l \right)^{\rd{A}\rd{A}'}{}_{\rd{B}\rd{B}'}$ where 
$U_l=(g_l^+,g_l^-)\in SO(4)$ and $g_l^{\pm}\in SU(2)$. Here we 
restrict ourselves to simple representations $(j,j)$.
For such representation matrices, the projection is defined by
\be
\pi\; :\; D^{(j,j)}(g_l^+,g_l^-)^{\rd{A}\rd{A}'}{}_{\rd{B}\rd{B}'}\longmapsto 
D^{(j,j)}(g_l,g_l)^{(\rd{A}\rd{A}')}{}_{(\rd{B}\rd{B}')}=D^{(2j)}(g_l)^{a}{}_{b}.
\ee
That is, we first restrict to a diagonal subgroup $\{(g,g)\} \subset \SO(4)$;
such a subgroup is isomorphic to $\SO(3)$.  The $(j,j)$
$\SO(4)$-representation matrix then becomes a $j\otimes j$
$\SU(2)$-representation matrix; from there we project to the highest
$\SU(2)$ irreducible, namely the $2j$ representation, as before.
The corresponding embedding of $\SO(3)$ spins into $\SO(4)$ spins
is then obviously $2j \mapsto (j,j)$.

Putting together the embeddings for intertwiners and spins, we obtain
the embedding (\ref{state_embed}) of $SO(3)$ LQG states into the
 $SO(4)$ boundary state
space of the model.

\item[(v)]

Lastly, with the embedding proposed in (iii,iv) above, and in light of
(\ref{area}),
the constraints (\ref{cas}) being used to solve the off-diagonal simplicity
constraints simply express the condition that three-dimensional
areas as determined by the SO(4) theory match three-dimensional
areas as determined by LQG.

\end{enumerate}
This concludes the discussion on the implementation of the constraints.
The last point to discuss is the dynamics.

\subsection{Dynamics}
\label{quantdyn_sect}

Following the strategy stated at the start of this section, consider
a triangulation formed by a single 4-simplex $v$. Denote the
boundary graph by $\Gamma_5$. A generic boundary state (satisfying
all kinematical constraints) is a function $\Psi(U_{ab})$, where
$a,b=1,...,5$, in the image $f[\Hil_{LQG}]$. We begin by writing the
transition amplitude between sharp values of the B's.  From this we
deduce the amplitude in terms of the U's, and then compute the
amplitude for the quantum state $\Psi(U_{ab})$. Beginning this
procedure,
\eqa
A(B_{ab})&=&\int \;\prod_a\; dV_a \;\prod_{(ab)}\;e^{iTr[B_{ab}V_a^{-1} V_b]}
 \nonumber \\
         &=&\int\;\prod_a\; dV_a\;\prod_{(ab)}dU_{ab}\; e^{iTr[B_{ab}U_{ab}]}\;
         \delta(V_a U_{ab} V_{b}^{-1})\nonumber \\
         &=&\int\;\prod_{(ab)}\; dU_{ab}\; e^{iTr[B_{ab}U_{ab}]}\int\;\prod_a\; 
         dV_a\; \delta(V_a U_{ab} V_{b}^{-1})
\neqa
so that the amplitude in the connection representation reads:
\be
A(U_{ab})=\int\;\prod_a\; dV_a\;\delta(V_a U_{ab} V_b^{-1}).
\label{ampliconn}
\ee
The integral is over a choice of $\SO(4)$ element $V$ at each node.
Let us define the bra
\be
\langle W|=\int\;\prod_{(ab)}\; dU_{ab}\; A(U_{ab})\langle U_{ab}|.
\ee
The quantum amplitude for a given state $|\psi\rangle$ in the boundary 
Hilbert space is:
\be A(\Psi):=\langle W|\Psi\rangle \label{ampli} \ee
As $\Psi$ satisfies all the kinematical constraints, it is of the form
$\Psi = f(\psi)$ for some $\psi \in \Hil_{LQG}$.  Let us consider
the specific case when $\psi$ is a spin-network state
$\psi = \psi_{\{j_{ab}\},\{i^a\}}$.
The amplitude is then given explicitly by
\be
A(\{j_{ab}\},\{i^a\}):=A(\psi_{\{j_{ab}\},\{i^a\}})= \sum_{i^a_+ i^a_-} \; 
15j\left(\left(\frac{j_{ab}}{2},\frac{j_{ab}}{2}\right);(i^a_+,i^a_-)\right)\;
 f^{i^a}_{i^a_+ i^a_-}
\label{ampliexp}
\ee
as can be easily seen by using (\ref{fi}) and decomposing \Ref{ampliconn} in
the $SO(4)$ spin-network basis.  This amplitude extends by linearity to
more general states $\Psi = f(\psi)$.

When we have a number of transitions between different 4-simplices, we have
to sum over boundary states around each, projecting each onto this
state $\langle W |$.
This gives transition amplitudes for boundary $SO(3)$ spin networks with
the amplitude generated by the partition function

\be
Z = \sum_{j_f, i_e}\ \prod_f (\dim {\scriptstyle\frac{j_f}{ 2}})^2\ \prod_v A(j_f,i_e).
\label{Z}
\ee

Observe that  the quantum dynamics defined above does not change
if we add the topological term (\ref{topo}) incorporating $\gamma$.
For, by changing the integration variable to
$\Pi_f := B_f + \frac{1}{\gamma} \star B_f$ in the path-integral,
the above derivation goes through in exactly the same manner, and the
final vertex amplitude differs at most by a constant, as determined
by the Jacobian of the transformation from $B$ to $\Pi$ 
(see appendix \ref{appsymp}). 

The problem is whether this quantum dynamics defines a non-trivial 
theory with general relativity as its low energy limit.  
A way of systematically exploring the low-energy behavior of a 
background independent quantum theory has recently been 
developed (after a long search; see for 
instance \cite{Iwasaki:1992qy}) in \cite{gravprop}. These techniques should shed some 
light on this problem.

\section{Conclusions}

We close with three observations. 

(i) There is a relation of the $\SO(4)$
states determined in this model to the projected spin network states
studied by Livine in \cite{etera}. (A similar approach is developed
by Alexandrov in \cite{sergei1}.) The constrained $\SO(4)$ states
that form the physical Hilbert space of the model presented here can
be constructed from (the euclidean analog of) these projected spin
networks. The Euclidean analog of the projected spin networks
defined in \cite{etera} are wavefunctions $\Psi[U_l, \chi_n]$
depending on an $\SO(4)$ group element for each link, and a vector
$\chi_n \in \SO(4)/\SO(3)$ at each node.  The wavefunctions are
labelled by an $\SO(4)$ representation $(j^+_l, j^-_l)$ for each
link, an $\SU(2)$ representation $j_{nl}$ for each node and link
based at that node, and an $\SU(2)$ intertwiner at each node. The
$\SO(4)$ spin-networks of the present paper can be obtained from
these projected spin networks by (i) setting $j^+_l = j^-_l \equiv
j_l$, (ii) setting $j_{nl} = j^+_l + j^-_l = 2 j_l$, and (iii)
averaging over $\chi_n$ at each node (concretely this averaging can
be done by acting on each $\chi_n$ with an $\SO(4)$ element $U_n$,
and then averaging over the $U_n$'s independently using the Haar
measure). Each of these three steps corresponds directly to solving
(i) the diagonal simplicity constraints, (ii) the off-diagonal
simplicity constraints, and (iii) the Gauss constraint. 

(ii) Livine and
Speziale have found an independent derivation of the vertex proposed
here  \cite{se}, based on the use of the coherent states they have
introduced in \cite{se1}.

(iii) The discreteness introduced by the Regge triangulation should
not be confused with the quantum-mechanical discreteness. The 
last is realized by the fact that variables that give a physical size to 
a Regge cell turn out to have discrete spectrum.   However, the two 
``discretenesses" end up to be related, because, due to the quantum 
discreteness,  in the quantum theory the continuum limit of the Regge 
triangulation turns out to be substantially different that the continuum 
limit of, say, lattice QCD or classical Regge calculus.
Intuitively, one can say that no further triangulation refinement can 
capture degrees of freedom below the Planck scale, since these do 
not exist in the theory.   The ultraviolet continuum limit is trivial in the 
quantum theory (for a discussion of this point, see \cite{libro}).  Thus, 
in the quantum theory the Regge 
triangulation receives, so to say \emph{a posteriori}, the physical interpretation
of a description of the fundamental discreteness of spacetime (the
one, say, that makes gravitational entropy finite \cite{Rovelli:1996ti}).
Numerous approaches to quantum gravity \emph{start} from the
assumption of a fundamental spacetime discreteness (\cite{discreetnes}); 
others use spacetime discreteness as a regularization
to be removed (\cite{loll});  the conventional formulation of loop quantum 
gravity \emph{derives} the discretization of spacetime from the quantization 
of continuum general relativity.  The derivation of loop quantum gravity given here 
is somewhat  intermediate: it starts from a lattice discretization, and 
later finds out that in the quantum theory, say, the area of the triangles
of the triangulation is discrete and therefore the regularization receives
a physical meaning. See also \cite{Zapata:2004xq}.  

In summary.  We have considered the Regge discretization of general relativity.
We have described it in terms of the Plebanski formulation of GR.
We have quantized the theory, on the basis of a flipped Poisson
structure and imposing the off-diagonal simplicity constraints
weakly.  This new way of imposing the simplicity constraints is weaker.
This weakening of the constraints is motivated by the observation
that they do not form a closed algebra, as well as by the realization
that a richer boundary space is needed for a correct classical
limit \cite{emanuele}. 

The theory we have obtained is characterized by the fact
that its boundary state space exactly matches that of ($\SO(3)$)
loop quantum gravity.  This can be seen as an independent derivation
of the loop quantum gravity kinematics, and, in particular, of the
fact that geometry is quantized.   A vertex amplitude has then been
derived from the discrete action, leading to a spin-foam model
giving transition amplitudes for loop quantum gravity states.

We expect that the model considered here will admit a group field
theory formulation \cite{gft} and that its vertex can be used to
generate the dynamics of loop quantum gravity also in the absence of
a fixed triangulation \cite{libro}. Whether this model is non-trivial and/or 
reproduces general relativity in an appropriate limit we do not know. 
Study of its semi-classical limit and $n$-point functions \cite{gravprop} 
should shed light on the discussion.

\centerline{------------------------------------}


Thanks to Alejandro Perez for help, criticisms and comments, to Sergei 
Alexandrov,  Etera Livine, Daniele Oriti and Simone Speziale for useful 
discussions. JE gratefully acknowledges support by an NSF International 
Research Fellowship under grant OISE-0601844

\appendix

\section{4-simplex geometry as gravitational field}
\label{app4s}

In the way we have defined the Regge triangulation above, the degrees of
freedom of the gravitational field are captured by the geometry of each
4-simplex.
Consider in each 4-simplex $v$ Cartesian coordinates such that one vertex of
the 4-simplex, say the vertex 5, has coordinates $(v_5)^a=0$, and the other
four
vertices have coordinates $(v_b)^a=\delta^a_b$. In these coordinates, the
metric
will not necessarily be $\delta_{ab}$, but rather take a form $g_{ab}(v)$.
The full
information about the geometry of the four simplex $v$ is then coded into the
ten quantities $g_{ab}(v)$.  Notice indeed that the shape of a 4-simplex in
$R^4$ is determined by 10 quantities (for instance the length of its ten
sides), and therefore the ten components of $g_{ab}(v)$ are the correct number
for capturing its degrees of freedom (possibly up to discrete degeneracies).

In particular, $g_{aa}$ (we will often drop the argument $(v)$ in the rest of
this appendix, since we deal here with a single 4-simplex) are (the square of)
the lengths of the four edges adjacent to the 5th vertex, and $g_{ab}$ are
essentially the angles between these edges.  (Also: $g^{aa}$ is the 
volume of the tetrahedron $a$, opposite to the vertex $a$, and $g^{ab}$ 
is the variable giving the angle between the normals to the tetrahedra 
$a$ and $b$.)

Thus, the discretized variables can be taken to be the ten
quantities $g_{ab}(v)$. This is of course no surprise at all, since
this is precisely a direct discretization of the variable
$g_{ab}(x)$ used by Einstein to describe the gravitational field,
here reinterpreted simply as a way to represent the geometry of each
elementary 4-simplex.

Another way of determining the geometry of a four simplex is to give
its ten areas. Let $A=1,2,3,4,5$ label the five tetrahedra bounding
the 4-simplex. The tetrahedron 1 is the one with vertices 2,3,4,5,
and so on cyclically.  Let $f_{AB}$ denote the triangle bounding the
two tetrahedra $A$ and $B$.  The triangle $f_{12}$ has vertices
$3,4,5$, and so on cyclically.   Let $A_{AB}\equiv A_{f_{AB}}$ be
the area of the triangle $f_{AB}$. Consider in particular the six
areas $A_{ab}\equiv A_{f_{ab}}$, of the six faces adjacent to the
vertex 5. A short computation, for instance using (\ref{area}),
shows that \be A_{ab}^2= g_{aa}g_{bb}-g_{ab}. \label{areagg} \ee
Define the angle variables $J_{aabc}\equiv J_{f_{ab}f_{ac}}$,
related to the angle between the triangles $f_{ab}$ and $f_{ac}$.
(We can also write $A^2_{ab}=J_{aabb}$.) Then again: \be J_{aabc} =
g_{aa}g_{bc}-g_{ab}g_{ac}. \label{arnglegg1} \ee The closure
relation of the bivectors of a tetrahedron implies \be A_{{15}}^2=
A_{{12}}^2+ A_{{13}}^2+
A_{{14}}^2+2(J_{{1123}}+J_{{1134}}+J_{{1142}}) \label{relations} \ee
and so on cyclically.  The six equations (\ref{areagg}) and the the
four equations (\ref{relations}) express the ten areas as functions
of the ten components of the metric.  Inverting these equations
gives the metric as a function of the areas: $g_{ab}(A_{AB})$, and
then, via  (\ref{arnglegg1}), the angles as functions of the areas:
$J_{{AABC}}(A_{AB})$. Therefore the full geometry of one 4-simplex
is determined by the ten areas $A_{AB}$.  Computing the two
functions $g_{ab}(A_{AB})$, and $J_{{AABC}}(A_{AB})$ explicitly
would be of great utility for quantum gravity.

\section{Existence of tetrad}
\label{app_tetrad}

In this appendix we show that the constraints (\ref{closurev},
\ref{aa}, \ref{bb}), at the classical discretized level, are
sufficient to imply the existence of a tetrad $e_{\mu}^I$ associated
with each tetrahedron $t$.  The main point is that the `dynamical
constraint' (\ref{ccv} or \ref{cc}) is \textit{not}
needed for the existence of $e_{\mu}^I(t)$.  Rather, as explained in
the main text, the role of (\ref{ccv}) or (\ref{cc}), or its
reformulation as (\ref{flat}), is to ensure that the geometries of
the tetrahedra in a single 4-simplex fit together to form a single
4-geometry.

The triad portion of the tetrad $e_{\mu}^I(t)$ associated with the
3-plane of the tetrahedron $t$ is uniquely determined (up to an
overall sign) by the $B(t)$'s of the faces of the tetrahedron. The
last component of the tetrad associated with the normal to $t$,
however, is not determined by the $B(t)$'s associated with $t$. This
final component is fixed only upon comparing with the tetrads at the
other tetrahedra in the 4-simplex via the parallel transport maps
$U_{tt'}(v)$.

Let us make all of this precise. Let $\Sigma$ denote the portion of
space-time associated with one of the 4-simplices $v$.
$\Sigma$ is a 4-manifold. Equip the tangent bundle of $\Sigma$ with a
background flat connection (so that we have a natural notion of
parallel translation between any two points, and a natural notion of
straightness).\footnote{
Within a single 4-simplex, or any non-closed chain of 4-simplices,
the choice of such a connection is a pure gauge choice.
}
In $v$, consider a tetrahedron $t$ with faces
labeled 1,2,3, and 4. For each face, we let $B_i^{\mu \nu}$ denote
the associated bivector living in the tangent space of $\Sigma$.
These bivectors satisfy
\begin{equation}
\label{bkgd_closure} \sum_{i=1}^4 B_i^{\mu\nu} = 0.
\end{equation}

What we wish to show in this appendix is that the discretized
constraints (\ref{closurev}, \ref{aa}, \ref{bb}) are sufficient to
imply that \textit{either}
\begin{equation}
\label{compatb} B_i^{IJ} = B_i^{\mu\nu} e^I_\mu e^J_\nu
\end{equation}
or
\begin{equation}
\label{compatbstar} {}^* B_i^{IJ} :=\half \epsilon^{IJ}{}_{KL}
B_i^{KL} = B_i^{\mu\nu} e^I_\mu e^J_\nu
\end{equation}
for some tetrad $e^I_{\mu}(t)$. We begin with a
\begin{lemma}
If $e^I_\mu$ is such that $B_i^{IJ} = B_i^{\mu\nu}e^I_\mu e^J_\nu$
for $i \in \{1,2,3\}$, and
\begin{equation}
\label{closure} \sum_{i=1}^4 B_i^{IJ} = 0,
\end{equation}
then $B_4^{IJ} = B_4^{\mu\nu}e^I_\mu e^J_\nu$ as well. An analogous
statement holds with $B_i^{IJ}$ replaced by ${}^*B_i^{IJ}$.
\end{lemma}
{\startproof

Using (\ref{closure}) and then (\ref{bkgd_closure}),
\begin{eqnarray}
B_4^{IJ} = -\sum_{i=1}^3 B_i^{IJ} = -\sum_{i=1}^3 B_i^{\mu\nu}
e^I_\mu e^J_\nu = B_4^{\mu\nu}e^I_\mu e^J_\nu .
\end{eqnarray}
The proof for the ${}^*B_i^{IJ}$ is similar. \finishproof}.

Thus, if we find a tetrad compatible with only $B_1^{IJ}, B_2^{IJ},
B_3^{IJ}$ (in the sense of (\ref{compatb}) or (\ref{compatbstar})),
then by imposing equation (\ref{closure}) (the closure constraint),
$B_4^{IJ}$ will then automatically be compatible with the same
tetrad, in the same sense.  In other words, by using the closure
constraint, we can ignore $B_4^{IJ}$.

Let $p$ denote the point in the tetrahedron shared in common by
faces 1,2, and 3. Let $v_1^\mu, v_2^\mu, v_3^\mu$ denote the three
edges emanating from this point, numbered such that face 1 is
between $v_2^\mu$ and $v_3^\mu$, etc, cyclically.  Then we can
choose our sign conventions such that
\begin{equation}
B_1^{\mu\nu} = 2v_2^{[\mu}v_3^{\nu]}, \; \text{and cyclically} .
\end{equation}
Specifying a tetrad $e^I_\mu$ is then equivalent to specifying the
three internal vectors
\begin{equation}
e_i^I:=v_i^\mu e^I_\mu, \; i \in \{1,2,3\},
\end{equation}
and the external unit normal $n_{\mu}$ to the tetrahedron. In terms
of the tetrad,
\begin{equation}
n_{\mu} := e^I_{\mu}n_I
\end{equation}
where $n_I$ is the unit normal to $\mspan\{v_1^{\mu}e_{\mu}^I,
v_2^{\mu}e_{\mu}^I, v_3^{\mu}e_{\mu}^I\}$. In terms of these
variables, the condition $B_i^{IJ} = B_i^{\mu\nu} e^I_\mu e^J_\nu$
for $i \in \{1,2,3\}$ becomes
\begin{equation}
B_1^{IJ} = 2 e_2^{[I}e_3^{J]}, \; \text{and cyclically} .
\end{equation}
and likewise the condition
${}^*B_i^{IJ}=B_i^{\mu\nu}e^I_{\mu}e^J_{\nu}$ becomes
\begin{equation}
{}^*B_1^{IJ} = 2 e_2^{[I}e_3^{J]}, \; \text{and cyclically} .
\end{equation}
We are now in a position to state the principal proposition and
sketch its proof.
\begin{proposition}
If $B_1^{IJ}, B_2^{IJ}, B_3^{IJ}$ are linearly independent and are
such that
\begin{equation}
\label{simplicity} {}^*B_i \cdot B_j = 0, \; i,j \in \{1,2,3\}
\end{equation}
then exactly one of the two following cases hold.
\begin{enumerate}
\item There exists $e_1^I, e_2^I, e_3^I$ such that
\begin{equation}
\label{beq} B_1^{IJ} = 2 e^{[I}_2 e^{J]}_3 \; \text{and cyclically},
\end{equation}
or
\item
there exists $e_1^I, e_2^I, e_3^I$ such that
\begin{equation}
\label{bstareq} {}^*B_1^{IJ} = 2 e^{[I}_2 e^{J]}_3 \; \text{and
cyclically} .
\end{equation}
\end{enumerate}
In either case, $(e^I_1, e^I_2, e^I_3)$ is unique up to $(e^I_1,
e^I_2, e^I_3) \mapsto (-e^I_1, -e^I_2, -e^I_3)$.
\end{proposition}
{\startproof

Equation (\ref{simplicity}), for the case $i=j$, using proposition
(3.5.35) of \cite{penroserindler}, tells us that there exists
$\alpha_i^I, \beta_i^I, \gamma_i^I, \delta_i^I$ such that
\begin{eqnarray}
B_i^{IJ} &=& 2 \alpha_i^{[I}\beta_i^{J]}
\\
{}^*B_i^{IJ} &=& 2 \gamma_i^{[I}\delta_i^{J]}
\end{eqnarray}
($i = 1,2,3$).  From the linear independence of the $B$'s, we know
in particular that none of them are zero. Therefore, if we define
for each $i$
\begin{eqnarray}
V_i &:=& \mspan\{\alpha_i, \beta_i\} \\
U_i &:=& \mspan\{\gamma_i, \delta_i\},
\end{eqnarray}
then $V_i$ and $U_i$ are each two dimensional.  $V_i$ is just the
two dimensional subspace uniquely determined by the bivector
$B_i^{IJ}$, whereas each $U_i$ is just the subspace uniquely
determined by the bivector ${}^*B_i^{IJ}$.

Let us now look at the rest of the equations (\ref{simplicity}). For
each $i\neq j$, they tell us that $\{\alpha_i, \beta_i, \alpha_j,
\beta_j\}$ and $\{\gamma_i, \delta_i, \gamma_j, \delta_j\}$ are each
linearly dependent.  $V_i$ and $V_j$ therefore non-trivially
intersect, as do $U_i$ and $U_j$, so that
\begin{eqnarray}
\dim (V_i \cap V_j) &>& 0 \\
\dim (U_i \cap U_j) &>& 0 .
\end{eqnarray}
But the linear independence of the $B$'s tells us that none of the
$V$'s can be the same, and none of the $U$'s can be the same. Thus,
for $i\neq j$,
\begin{eqnarray}
\dim (V_i \cap V_j) &<& 2 \\
\dim (U_i \cap U_j) &<& 2 ,
\end{eqnarray}
whence
\begin{eqnarray}
\dim (V_i \cap V_j) &=& 1 \\
\dim (U_i \cap U_j) &=& 1 .
\end{eqnarray}
Let $f_1^I$ be any non-zero vector in $V_2 \cap V_3$, etc
cyclically. Likewise let $\tilde{f}_1^I$ be any non-zero vector in
$V_2 \cap V_3$, etc cyclically. One can prove that exactly one of
$\mspan\{f_1, f_2, f_3\}$ and
$\mspan\{\tilde{f}_1,\tilde{f}_2,\tilde{f}_3\}$ is three
dimensional, and the other is one dimensional. To prove this is
non-trivial; we leave it as an exercise to the reader. In proving
this, one uses in a key way both the assumption that the $B$'s are
linearly independent and the fact that for each $i$, $V_i$ and $U_i$
are orthogonal complements (which one can also show).

In the case where $\{f_1, f_2, f_3\}$ is linearly independent, by
setting $e_i = \lambda_i f_i$, one can solve for $\lambda_i$ such
that (\ref{beq}) holds.  These $\lambda_i$ are unique up to
$(\lambda_1,\lambda_2,\lambda_3)\mapsto (-\lambda_1, -\lambda_2,
-\lambda_3)$, so that $e_i$ is unique up to $(e_1,e_2,e_3) \mapsto
(-e_1,-e_2,-e_3)$. Furthermore, one can show the fact
$\dim(\mspan\{\tilde{f}_1,\tilde{f}_2, \tilde{f}_3\})=1$ implies
that there exists no $e_i$ such that (\ref{bstareq}) holds.

In the case where $\dim(\mspan\{f_1, f_2, f_3\})=1$ and
$\{\tilde{f}_1, \tilde{f}_2, \tilde{f}_3\}$ is linearly independent,
the situations are obviously reversed.  There exist no $e_i$ such
that (\ref{beq}) holds, whereas there exist $e_i$'s unique up to
$(e_1,e_2,e_3)\mapsto(-e_1, -e_2, -e_3)$, such that (\ref{bstareq})
holds.

Thus we have the proposition.    \finishproof}.  

\section{Symplectic structure}
\label{appsymp}

The action is
\begin{equation}
\label{disc_action} S = \half \sum_{f \in int\Delta} \Tr (B_f(t) U_f(t))
+
\half \sum_{f \in \partial\Delta} \Tr (B_f U_f R_f) .
\end{equation}

In the case when one fixes the $B$ variables on the boundary, one
sets $R_f =$ identity, as then the boundary term reduces to the
usual classical boundary term appropriate when fixing the $B$
variables on the boundary \cite{danieleboundary}.  When one fixes
the $U$'s (i.e., the connection) on the boundary, analogy with the
classical theory suggests that there should be no boundary term.
However, in a discrete theory this is not literally possible, as
then no boundary variables would appear in the action at all. For
the case of fixing the $U$'s on the boundary, we therefore suggest
the following prescription.  We want the terms in the boundary sum
to essentially be the same as the terms in the interior sum: this
will be the case if $R_f$ is the holonomy around the rest of the
link of $f$ outside of $\Delta$.  Of course, the problem is that
this part of the holonomy around $f$ is not determined by any
dynamical variables. We propose the following prescription. Whenever
varying the action, keep the $R_f$'s fixed; however, whenever
evaluating anything on extrema of the action, set $R_f$ equal to
what we know it `should' be: namely, the holonomy along the part of
the link of $f$ outside of $T$ as determined by the BF equations of
motion. The BF equations of motion dictate that all holonomies
around closed loops are trivial. Thus, our prescription is, whenever
evaluating an expression onshell, set $R_f = U_f^{-1}$. A remark is
in order as to why the BF equations of motion are used. The BF
equations of motion are the equations of motion coming from the
unconstrained action (\ref{disc_action}). We use the unconstrained
action in deriving the symplectic structure because our general
philosophy is that we are quantizing GR as a BF theory
\textit{constrained at the quantum level}.  That is, we first
compute the symplectic structure and quantize everything as though
it were pure BF theory. Only then, after the `kinematical'
quantization is complete, do we impose the simplicity constraints
and obtain GR.

Let us determine the symplectic structure from the action, when we
fix the $U$'s on the boundary.  We will use a method essentially
coming from that described in \cite{asht_et}, and briefly mentioned
in \cite{alrev}; the method is the following. We first apply an
arbitrary variation $\delta$ to the action without fixing any
variables on the boundary; then we restrict to solutions of the
equations of motion. The result will be a boundary term
$\underline{\Theta}(\delta)$ depending linearly on the variation.
This gives us a linear map from variations (at the space of
solutions) into real numbers. This linear map is then extended
off-shell basically by keeping the same expression in terms of the
basic variables.\footnote{If there are constraints, this expression
has ambiguities.  One uses the ``obvious'', simplest expression in
this case.}
We thereby obtain a linear map $\Theta$ from arbitrary variations of
boundary data into real numbers --- that is, a \textit{one-form} on
the space of boundary data. This $\Theta$ is the canonical one form
on the boundary phase space. The symplectic structure is then the
exterior derivative $\Omega = \dbldif \Theta$.

Let us apply this procedure. Varying the action (\ref{disc_action})
gives
\begin{eqnarray}
\nonumber \delta S &=& \half \sum_{f \in int\Delta} \Tr ((\delta
B_f) U_f(t))
+ \half \sum_{f \in int\Delta} \Tr (B_f(t)(\delta U_f(t))) \\
&+& \half \sum_{f \in \partial{\Delta}} \Tr ((\delta B_f) U_f R_f)
+ \half \sum_{f \in \partial{\Delta}} \Tr (B_f (\delta U_f)R_f) .
\end{eqnarray}
Next, restrict to the case of $B$'s and $U$'s that satisfy the
equations of motion. Then, when $\delta U_f$ on the boundary is zero
for $f \in \partial \Delta$, the entire expression above must vanish. But
only the fourth term depends on $\delta U_f$ for $f \in
\partial \Delta$.  Therefore, the other terms must vanish (on solutions)
no matter what is $\delta U$ on the boundary. Therefore, on
solutions, at most the fourth term above can survive, and we have
\begin{equation}
\delta S = \half \sum_{f \in \partial{\Delta}} \Tr (B_f (\delta
U_f)R_f) .
\end{equation}
Applying the equations of motion to $R_f$, we finally set $R_f =
U_f^{-1}$, giving us
\begin{equation}
\underline{\Theta}(\delta) = \half \sum_{f \in
\partial{\Delta}} \Tr (B_f(\delta U_f)U_f^{-1}) .
\end{equation}
Extending $\underline{\Theta}$ to arbitrary variations of the
boundary data, keeping the same expression in terms of the basic
variables, we obtain the canonical one-form
\begin{equation}
\label{canoneform}\Theta(\delta) = \half \sum_{f \in
\partial{\Delta}} \Tr (B_f(\delta U_f)U_f^{-1}) .
\end{equation}

We next claim that $\big[(\delta U_f)U_f^{-1}\big]^I{}_J$ is an
element of $\so(4)$. To show this, let $U_f(\lambda)$ denote a
one-parameter path in $\SO(4)$ such that $\delta$ is the tangent to
the path at $\lambda = 0$. Then
\begin{equation}
(\delta U_f)U_f^{-1} = \deriv{}{s} U_f(s) |_{s=0} U_f(0)^{-1} =
\deriv{}{s} [U_f(s) U_f(0)^{-1}]_{s=0}
\end{equation}
The argument in brackets is the identity at $s=0$, so that one has
the action of a tangent vector at the identity on the matrix
elements of the group. Thus one has an element of the Lie algebra.

Introduce a basis $\{\xi_\alpha\}_{\alpha=1,\dots 6}$ of $\so(4)$,
orthonormal with respect to the trace --- i.e. such that
$\Tr(\xi_\alpha \xi_\beta) = -\xi_\alpha^{IJ} \xi_{\beta IJ}
= -\delta_{\alpha \beta}$. Denote the
components of a Lie algebra element with respect to this basis just
by adding a $\alpha, \beta, \gamma \dots$ superscript or
subscript. (\ref{canoneform}) can then be written
\begin{equation}
\Theta(\delta) = -\half \sum_{f \in \partial{\Delta}} (B_f)_\alpha
\big[(\delta U_f)U_f^{-1}\big]^\alpha .
\end{equation}
Define
\begin{equation}
\mu_f(\delta)^\alpha := \big[(\delta U_f)U_f^{-1}\big]^\alpha
\end{equation}
so that $\mu_f^\alpha$ is a one-form on the copy of $\SO(4)$
parametrized by $U_f$.  We can then write the canonical one form
$\Theta$ as
\begin{equation}
\Theta = -\half \sum_{f \in \partial{\Delta}} (B_f)_\alpha
\mu_f^\alpha .
\end{equation}
For each face $f$, and each basis element $\xi^\alpha \in \so(4)$, let
$\delta^R_{f,\alpha}$ denote the corresponding right invariant
vector field on the associated copy of $\SO(4)$. We then have
\begin{equation}
\mu_f(\delta^R_{f,\alpha})^\beta = \delta^\beta_\alpha .
\end{equation}
Thus the $\mu_f^\alpha$ form the basis of right invariant
\textit{one-forms} dual to the basis $\delta^R_{f,\alpha}$ of right
invariant \textit{vector fields}.  These right invariant one forms
have been well studied, for example, by Cartan \cite{cartan}. They
satisfy
\begin{equation}
\dbldif \mu_f^\alpha = - \half f^\alpha{}_{\beta \gamma} \mu_f^\beta
\dblwedge \mu_f^\gamma
\end{equation}
where $f^\alpha{}_{\beta \gamma}$ are the structure constants of the
Lie algebra.  Using this identity to compute the symplectic
structure from $\Theta$, we obtain
\begin{equation}
\Omega := \dbldif \Theta = - \half \sum_{f \in \partial{\Delta}}
(\dbldif (B_f)_\alpha) \dblwedge \mu_f^\alpha + \frac{1}{4} \sum_{f
\in
\partial{\Delta}} f^\alpha{}_{\beta \gamma} (B_f)_\alpha \mu_f^\beta
\dblwedge \mu_f^\gamma .
\end{equation}
From this, one can read off the Poisson brackets. As there are no
$\dbldif B \wedge \dbldif B$ terms, the $U_f$'s commute. The first
term above gives the Poisson bracket between the $B$'s and the
$U$'s, and the second gives the Poisson brackets among the $B$'s.
Explicitly,
\begin{eqnarray}
\{U_f, U_{f'} \} &=& 0 \\
\{(B_f)_\alpha, U_{f'} \} &=& 2 \delta_{f,f'} \xi_\alpha U_{f'} \\
\{(B_f)_\alpha, (B_{f'})_\beta \} &=& 2 \delta_{f,f'}
f^\gamma{}_{\alpha \beta} (B_f)_\gamma .
\end{eqnarray}
The second two equations can be summarized by stating that the
action generated by $(B_f)_\alpha$ on $U_f$ is that of the right
invariant vector field field determined by $2 \xi_\alpha \in
\so(4)$. The above Poisson brackets can also be more explicitly
written in terms of matrix elements
\begin{eqnarray}
\{(U_f)^I{}_J, (U_{f'})^K{}_L \} &=& 0 \\
\{B_f^{IJ}, (U_{f'})^K{}_L\} &=& 2 \delta_{f,f'}
\delta^{K[I}(U_f)^{J]}{}_L =
\delta_{f,f'}\left(\delta^{KI}(U_f)^J{}_L
- \delta^{KJ}(U_f)^I{}_L\right)\\
\{(B_f^{IJ}, B_{f'}^{KL} \} &=&
4\delta_{f,f'}\delta_M{}^{[I}\delta^{J][K} \delta^{L]}{}_N B_f^{MN}
.
\end{eqnarray}
If we define $(\tau^{IJ})^M{}_N := 2 \delta^{M[I}\delta^{J]}{}_N$,
this becomes (\ref{pbUU}).

Thus we obtain the symplectic structure from the action when we fix
the $U$'s on the boundary.  The case of fixing the $B$'s on the boundary
is not as well-understood.  Application of the above prescription to
this case seems to lead to a non-$\SO(4)$-gauge-invariant symplectic structure
in which all the $B$'s commute.  For the present, we simply do not address this
problem, and take the symplectic structure to be the one determined with $U$'s fixed.

So far we have seen that the action we have considered gives the unflipped
Poisson structure.

Recall, however, that in the LQG approach
the action that is quantized is the GR one with a topological term
that doesn't change the equations of motion:
\be S_{LQG}=\int\;*(e\wedge e)\wedge F +\frac{1}{\gamma}\int\;
(e\wedge e)\wedge F \ee where $\gamma$ is the so called Immirzi
parameter. Recall that the introduction of this topological term is
\textit{required} in order to have a theory of connections on the
boundary: without it, as shown by Ashtekar, the connection variable
does not survive the Legendre transform \cite{abhay1991}. Let us
therefore consider the discretization of the modified BF action:
\be S=\int \left(B+\frac{1}{\gamma}{}^*B\right)\wedge F +
\phi_{IJKL}B^{IJ}\wedge B^{KL} \label{bfholst} \ee and follow the
lines presented in this paper. Define then
$\Pi_f:=B_f+{\scriptstyle \frac{1}{\gamma}}*B_f$;
the symplectic
structure is then such that, for each face, the $\Pi_f^{IJ}$ are
identified with the generators of $\SO(4)$. The constraints $C_{ff}$
imply, in terms of these new variables:
\be
\left(1+\frac{1}{\gamma^2}\right)\; {}^*\Pi_f\cdot \Pi_f-\frac{2}{\gamma}\;
\Pi_f\cdot \Pi_f=0
\label{sigmaff}
\ee
Note that for $\gamma \ll  1$ and $\gamma \gg  1$ one recovers the
same constraint ${}^*\Pi_f\cdot \Pi_f \sim 0$ being used before,
whence, in these two limits, the diagonal simplicity constraints
again imply simplicity of the representations of the edges ($j^+ =
j^-$). (Thanks to Alejandro Perez for pointing this out.)
Furthermore, as mentioned in \S\ref{quantdyn_sect}, neither does the
dynamics change with the introduction of the $\gamma$-term in the
action. What is different, then, with the introduction of the
$\gamma$ term, when $\gamma \ll 1$ or $\gamma \gg  1$?  In the
$\gamma \ll  1$ case, the symplectic structure in terms of $B$ is
different --- it is the flipped symplectic structure. Is there an
independent reason to prefer $\gamma \ll  1$ over $\gamma \gg  1$?
Yes: the latter is unphysical, because it leads to a macroscopic
area spectrum in the boundary theory.  Thus, when we look at things
more carefully, we do in fact see that the flipped symplectic
structure is a consistent choice. And with this flipped symplectic
structure, as noted in the main text, the off-diagonal simplicity
constraints, in the form (\ref{cas}), lead precisely to the model
presented in this paper.

As a side remark, consider the case $\gamma \gg  1$.  In this case,
one obtains the more standard symplectic structure (\ref{sympst}).
As noted in \S\ref{intspace_sect}, this symplectic structure, along
with the constraint (\ref{cas}), imposed as discussed above, leads
to the Barrett-Crane model.  Thus the $\gamma \gg  1$ case, which we
argue against, can be viewed as yielding the Barrett-Crane model.

\end{document}